\documentclass[journal]{IEEEtran}
\usepackage{amsfonts}
\usepackage{indentfirst}
\usepackage{amsmath}
\usepackage{amsthm}
\usepackage{amssymb}
\usepackage{color}
\usepackage{graphicx}
\usepackage[para]{threeparttable}
\usepackage{booktabs}
\usepackage{makecell}
\usepackage{diagbox}
\usepackage{multirow}
\usepackage{epstopdf}
\usepackage{mathrsfs}%花写字母
\usepackage{bm}%加粗公式
\usepackage{subfigure}%图片子图
\usepackage{multicol}%同一页中实现不同分栏
\usepackage{stfloats}

\theoremstyle{remark} % ±àºÅÏÔÊ¾·ç¸ñ£¬¿ÉÑ¡ plain or definition or remark

\newtheorem{Lemma}{\textit{Lemma}}
\newtheorem{Remark}{\textit{Remark}}

\usepackage[noend]{algpseudocode}
\usepackage{algorithmicx,algorithm}
\algdef{SE}[DOWHILE]{Do}{doWhile}{\algorithmicdo}[1]{\algorithmicwhile\ #1}%伪代码
\allowdisplaybreaks[4]

\ifCLASSINFOpdf
\else
\fi
\begin{document}
	\title{Waveform and Filter Design for Integrated Sensing and Communication Against Signal-dependent Modulated Jamming}
	\author{Yu Zhou, $ \text{Qiao Shi}^{*} $, \IEEEmembership{Member, IEEE}, Zhengchun Zhou, \IEEEmembership{Member, IEEE},\\ Zilong Liu, \IEEEmembership{Senior Member, IEEE}, and Pingzhi Fan, \IEEEmembership{Fellow, IEEE}

	\thanks{This work was supported in part by the National Natural Science Foundation of China under Grants U23A20274, 62350610267, and 62131016, in part by the Natural Science Foundation of Sichuan Province under Grants 2024NSFTD0015 and 2024NSFSC1418, in part by project of central government to guide local scientific and technological development under Grant 2024ZYD0012, in part by the Open Research Fund of Key Laboratory of Analytical Mathematics and Applications (Fujian Normal University), Ministry of Education, under Grant JAM2406, in part by the China Postdoctoral Science Foundation under Grant 2023M742901. The work of Z. Liu was supported in part by the UK Engineering and Physical Sciences Research Council under Grant EP/Y000986/1 (`SORT').}
	\thanks{Yu Zhou, Qiao Shi and Zhengchun Zhou are with the School of Information Science and Technology, Southwest Jiaotong University, Chengdu 611756, China (e-mail: 2021201698@my.swjtu.edu.cn; qiaoshi@swjtu.edu.cn; zzc@swjtu.edu.cn). (\emph{Corresponding author: Qiao Shi}.)}
	\thanks{Zilong Liu is with the School of Computer Science and Electronics Engineering, University of Essex, Colchester CO4 3SQ, U.K. (e-mail: zilong.liu@essex.ac.uk).}
	\thanks{Pingzhi Fan is with the Key Laboratory of Information Coding and Wireless Communications, Southwest
	Jiaotong University, Chengdu 611756, China (e-mail: pzfan@swjtu.edu.cn).}}

	%\markboth{Journal of \LaTeX\ Class Files,~Vol.~14, No.~8, August~2015}

	\maketitle
	
	\begin{abstract}
		This paper focuses on an integrated sensing and communication (ISAC) system in the presence of signal-dependent modulated jamming (SDMJ). Our goal is to suppress jamming while carrying out simultaneous communications and sensing.
		We minimize the integrated sidelobe level (ISL) of the mismatch filter output for the transmitted waveform and the integrated level (IL) of the mismatch filter output for the jamming, under the constraints of the loss in-processing gain (LPG) and the peak-to-average power ratio (PAPR) of the transmitted waveform.
		Meanwhile, the similarity constraint is introduced for information-bearing transmit waveform.
		We develop a decoupled majorization minimization (DMM) algorithm to solve the proposed multi-constrained optimization problem. In contrast to the existing approaches, the proposed algorithm transforms the difficult optimization problem involving two variables into two parallel sub-problems with one variable, thus significantly speeding up the convergence rate. 
		Furthermore, fast Fourier transform (FFT) is introduced to compute the closed-form solution of each sub-problem, giving rise to a greatly reduced computation complexity.
		Simulation results demonstrate the capabilities of the proposed ISAC system which strikes a proper trade-off among sensing and jamming suppression. 
	\end{abstract}
	

	\IEEEpeerreviewmaketitle
	\section{Introduction}
	\subsection{Background and Related Work}
	\IEEEPARstart{T}{he} ever-growing wireless devices and digital applications across the world have led to increasingly congested spectrum. Against this problem, radar sensing and communication spectrum sharing (RCSS) has been extensively studied in recent years, where the goal is to operate both functionalities simultaneously over the same frequency bands \cite{L. Zheng-2019}, \cite{S. Lu-2024}. There are two major research directions for RCSS \cite{S. Lu-2024}, \cite{M. Liu-2023}: 1) radar-communication coexistence (RCC); 2) integrated sensing and communication (ISAC).\par
	
	In RCC, the radar and communication systems share the same spectrum, but the transmitted signals for these two are designed independently. Because of this, cross-interference suppression \cite{M. Bica-2019} is needed to avoid potential degradation of their performance \cite{B. Li-2017}, \cite{F. Liu-2018-1}. Also, real-time collaboration between the two systems is required, yet at the expense of increased system complexity and communication overhead. On the other hand, ISAC aims at designing an integrated system by simultaneously performing sensing and communication tasks at the same hardware platform.
	Therefore, compared with RCC, ISAC may provide considerable gains in terms
	of spectral/energy/hardware efficiencies as well as significant mutual enhancement of the two functionalities \cite{F. Liu-2022}-\cite{F. Dong-2023-resource}. From the waveform standpoint, there are three main types of ISAC waveform design schemes \cite{X. X. Yu-2022}: 1) radar waveform-based \cite{G. N. Saddik-2007}, \cite{D. Gaglione-2018}; 2) communication waveform-based \cite{J. Johnston-2022}, \cite{M. F. Keskin-2021}; 3) joint waveform design \cite{J. Yang-2020}-\cite{Y. Dong-2023-letter}.\par
	
	The radar waveform-based scheme was advocated in the early stage of ISAC research, where the communication information is embedded by modifying traditional radar waveforms or utilizing the index modulation (IM) technique \cite{A. Hassanien-2016}. For example, one can send digital data by applying phase modulation in frequency modulated continuous-wave (FMCW) and linear frequency modulation waveforms \cite{C. Sahin-2017}, \cite{G. N. Saddik-2007}. However, the radar waveform-based scheme mainly uses inter-pulse modulation to embed communication information. Thus, the achievable communication data rate is limited and may not be able to support many real-world communication needs. \par
	
	To achieve a superior communication functionality, various communication waveform-based schemes have been studied, whereby communication waveforms directly act as a radar probing signal to locate, detect and track targets \cite{D. Ma-2020}. In particular, as a commonly used waveform in modern communication systems, orthogonal frequency division multiplexing (OFDM) has been widely exploited as ISAC waveforms \cite{C. Sturm-2011}, \cite{Y. Liu-2017}.
	However, OFDM suffers from the high peak-to-average power ratio (PAPR) problem, resulting in distorted ISAC waveforms as well as limited communication and sensing ranges. In \cite{Y. Huang-2022}, a new waveform design algorithm is proposed to reduce PAPR in OFDM-based ISAC systems.
	Furthermore, due to the randomness of communication data, it is difficult to deliver guaranteed sensing performance.\par
	
	The joint ISAC waveform design is more flexible than the aforementioned two schemes, where an integrated waveform is developed to deliver both the sensing and communication functionalities. Many research attempts have been made to study joint ISAC waveform design \cite{F. Liu-2018-2}-\cite{C. G. Tsinos-2021}. In \cite{F. Liu-2018-2}, ISAC waveform is directly designed by minimizing the multi-user interference (MUI), in which both the omnidirectional and directional beam pattern design problems are considered. To further improve the radar beam pattern performance, the desired radar beam pattern is realized by imposing constraints on the covariance matrix of the transmitted signal of each antenna \cite{F. Liu-2018-3}. In \cite{X. Liu-2020}, the individual precoders of the radar and communication are designed to
	optimize the performance of the multi-input multi-output (MIMO) radar beamforming
	while meeting the signal-to-interference-noise ratio (SINR) constraint for the users.
	In \cite{F. Dong-2023}, a sophisticated beamforming design for multi-user ISAC
	systems is proposed by additionally taking the physical layer security into account.
	Albeit these works have studied the transmitted waveform design, the joint design of the transmitted waveform and the receive filter has not been taken into account. To make a difference, joint transceiver beamforming designs for MIMO radar and multi-user communications are studied in \cite{L. Chen-2022} and \cite{C. G. Tsinos-2021}, respectively.\par
	
	\subsection{Motivations}
	Note that the aforementioned studies mainly focus on improving the performance of sensing and communication by minimizing the MUI, maxmizing the SINR, and so on. An overly simple sensing environment without jamming is often assumed. However, with the tremendous advances of the digital and electronic technologies, future ISAC systems are facing an increasingly complex environment, whereby the interference consists of not only clutter, but also intentional jamming \cite{S.D. Berger-2003}.  For example, the digital radio frequency memory (DRFM) jammer, which can store, copy and forward radar signals, brings serious challenges to radar systems \cite{S.D. Berger-2003}. In this case, the target detection performance of the aforementioned designs may not satisfy the practical requirements.\par

	It is worth mentioning that the radar waveform design problems for suppressing DRFM jamming have been well studied independently \cite{W. Xiong-2017}-\cite{S. Wei-2024}. There are two types of DRFM jamming: full pulse repetitive jamming and signal-dependent modulated jamming (SDMJ). In comparison, the latter is more interesting because it only samples and forwards part of the radar pulse, thus giving rise to a superior jamming efficiency \cite{S. Hanbali-2019}.
	In \cite{W. Xiong-2017} and \cite{J. Chen-2019}, the time-frequency distribution of the SDMJ is studied, in which a filter is designed to suppress jamming in the time-frequency domain. In \cite{S. Lu-2018}, integrated cross-correlation energy between the transmitted waveform and the jamming signal is minimized to combat the jamming. In \cite{K. Zhou-2020}, the transmitted waveform and the mismatch filter are jointly designed in the presence of the SDMJ. The major drawback of their scheme is its high computation complexity. To address this problem, a fast algorithm for designing the waveform and filter is introduced to suppress jamming  \cite{K. Zhou-2022}. Subsequently, various methodologies are proposed to improve the performance of DRFM jamming suppression \cite{M. Ge-2021}, \cite{S. Wei-2024}.

	It is noted that the existing waveform design methods for radar-only systems cannot be directly applied to ISAC under jamming. Motivated by this, we propose a transmit waveform and receive filter design by considering the  SDMJ, which leads to novel transmitted waveforms that can be exploited for improving sensing and communication. Some preliminary results have been published in \cite{Y. Zhou-2023}, where only phase difference constraint is considered to realize the communication.
	In this paper, we shall provide a more complete and detailed theoretical and numerical study. 
	\subsection{Contributions }
	Specifically, the main contributions of this work are summarized as follows.
	\begin{itemize}
		
		\item  An important scenario where target sensing and communication are simultaneously achieved in the presence of the SDMJ is considered. We formulate an optimization problem to jointly design the ISAC waveform and the filter under the constraints of the loss in-processing gain (LPG\footnote{The LPG is defined as the ratio between the radar SNR gathered with mismatched filter and  maximum radar SNR obtained by the matched filter.}), PAPR, and energy. Our core idea is to minimize the integrated sidelobe level (ISL) of the mismatch filter output for ISAC waveform and the integrated level (IL) of the mismatch filter output for jamming signal. Furthermore, we consider the waveform similarity constraint\footnote{The similarity constraint forces the designed waveform to be close to a reference waveform with the desirable properties \cite{J. Li-2006}.} to ensure communication performance by introducing a penalty term. By adjusting the penalty parameters, one can strike a flexible trade-off between sensing, jamming suppression and communication performance.\par
		
		\item  A new low-complexity decoupled majorization minimization (DMM) algorithm is developed to solve the  formulated non-convex and NP-hard problem. Unlike the existing methods that generally solve the above joint design problem by using the alternating optimization based majorization minimization (MM) algorithm \cite{K. Zhou-2022}, \cite{K. Zhou-2022-J}, we transform the proposed optimization problem involving two variables into two parallel sub-problems with one variable, thus significantly speeding up the convergence rate.
		Meanwhile, fast Fourier transform (FFT) is introduced to compute the closed-form solution of each sub-problem. Compared to the alternating optimization based MM (AMM) algorithm, the proposed DMM algorithm offers a significant improvement in computational efficiency.
		
		\item To analyze the superiority of the proposed DMM algorithm more effectively, the convergence and the computational complexity of the DMM algorithm is further analyzed.
		Finally, simulation results are presented to validate the excellent performance of the proposed method for partial pulse repeater jamming (PPRJ) and repetitive repeater jamming (RRJ) scenarios. 
		Specifically, the sensing and communication performance of  the proposed joint design method outperforms the conventional ISAC waveform in the presence of the SDMJ.
		
	\end{itemize}\par
	%\subsection{Organization}
	The rest of the paper is organized as follows. Section II describes the system model of the ISAC system in the presence of jamming. In Section III, we first propose a joint optimization problem of the ISAC waveform and the filter, and deduce the solution procedures of the DMM algorithm. We also analyze the computational complexity and the convergence of the DMM algorithm. Section IV presents the simulation results and Section V concludes this work.\par
	
	%\subsection{Notations}
	\emph{Notations}:
	Matrices are denoted by bold uppercase letters and vectors
	are denoted by bold lowercase letters. $ \mathbf{I} $ and $ \mathbf{O} $ respectively denote the identity and zero matrix, with the size determined by a subscript or from the context. Superscripts $ (\cdot)^{*} $, $ (\cdot)^{T} $ , $ (\cdot)^{H} $ denote the complex conjugate, transpose, conjugate transpose, respectively. $\mathbf{F}=(F_{m,n})$ represents the discrete Fourier matrix with size $2L\times2L $, and $ F_{m,n}=e^{-j\frac{2mn\pi}{2L}},0\leq m,n< 2L $. $\mathbb{C}^{L} $ and $\mathbb{C}^{L\times L} $ denote the $ L $-dimensional complex vector and the $ L\times L $-dimensional complex matrix spaces, respectively. $ \text{dist}(\cdot) $ indicates the distance function. $ |x| $ and $ \parallel\mathbf{x}\parallel_{2} $ denote the modulus of $ x $ and the $ l_{2} $ norm of the vector $ \mathbf{x} $, respectively. $ \text{Re}\{\cdot\}$ and $ \text{arg}(\cdot) $  denote the real part and the phase of a complex number, respectively. The symbol $ \lfloor \cdot \rfloor $ and $ \circ $ denote the floor operation and the Hadamard product, respectively. $ \mathbf{Tr}(\cdot) $ indicates the trace of a square matrix. $ \lambda_{\max}(\mathbf{X}) $ denotes the largest eigenvalue of $ \mathbf{X} $. $ \text{Diag}(\mathbf{x}) $ 
	is a diagonal matrix formed with $ \mathbf{x} $ as its principal diagonal. $ \text{vec}(\cdot) $ is the vectorization operator.

	\section{System Model}
\begin{figure*}[h]
		\centering
		\includegraphics[width=0.9\linewidth]{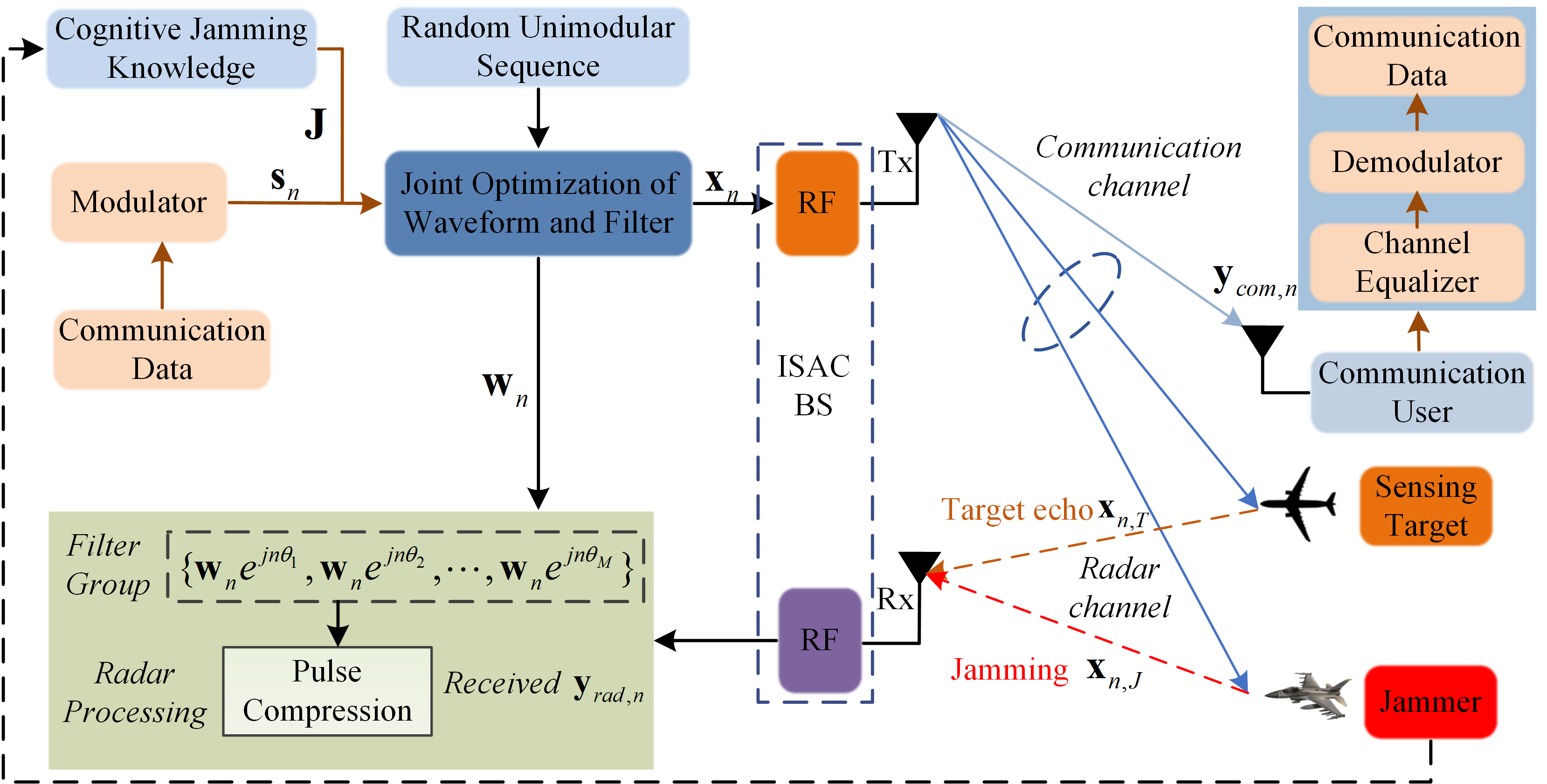}
		\caption{The signal processing block diagram of the proposed ISAC system.}
		\label{fig:system_structure}
	\end{figure*}

	We consider a single carrier ISAC system, where  the base station (BS) transmits a waveform with multiple pulses for sensing a point  target and serving a single-antenna communication user, as shown in Fig. \ref{fig:system_structure}\footnote{This paper considers a mainlobe DRFM forwarding jamming \cite{S.D. Berger-2003}, which has strong directivity and is usually directly sent toward radar receiver, thus greatly restricting the radar detection ability because of its strong energy and similarity with the target echo in space, time and frequency dimensions. Based on this, we assume that the user will not be affected by the jamming.}.
	The DRFM jammer intercepts the transmitted ISAC waveform and carries out a series of operations, including sampling, storage, and signal reconstruction, to disturb the sensing system.\par

	We assume that $ N $ ISAC pulses are transmitted within a coherent processing interval (CPI) with a constant pulse
	repetition frequency (PRF) $ f_{r} $, and $ \mathbf{x}_{n}=[x_{n,0},\cdots,x_{n,L-1}]^{T}\in\mathbb{C}^{L}$ denotes the discrete baseband signal transmitted within the $ n $-th constant pulse repetition time (PRT) $ T_{r} $, where $ T_{r}=1/f_{r} $ and $ L $ is the discrete length of the baseband signal. Then, the transmitted signal of the $ n $-th pulse can be expressed as
	\begin{equation}\label{key}
		x_{n}(t)=\sum_{l=0}^{L-1}x_{n,l}\Omega(t-lT_{c}), n=0,1,\cdots,N-1, 
	\end{equation}
	where  $ T_{c} $ represents a code duration, and $ \Omega(t) $ denotes a unit energy baseband shaping pulse, i.e., $ \int_{-T_{c}/2}^{T_{c}/2}|\Omega(t)|^{2}dt=1 $.
	Then, on the one hand, we receive and further demodulate $x_{n}(t) $ at the communication user end.
	On the other hand, the ISAC system receives the signal $ y_{rad,n}(t) $ and carries out target sensing by coherent processing, where $ y_{rad,n}(t) $ is the summation of the jamming, target echo and noise, which can be expressed as
	\begin{equation}\label{key}
		\begin{split}
			y_{rad,n}(t)=&\alpha_{n,T}x_{n}(t-t_{0})e^{j2\pi f_{d}nT_{r}}\\
			&+\alpha_{n,J}x_{n,J}(t-t_{J})+v_{rad,n}(t),
		\end{split}
	\end{equation}
	where $\alpha_{n,T}$ and $\alpha_{n,J}$ respectively represent the responses of the target and the jamming,  $ t_{0}$ and $ t_{J} $ denote the corresponding time delay of the target and the jamming, respectively, 
	$ \theta=2\pi f_{d}T_{r} $ denotes the normalized Doppler shift, $ f_{d}$ is the Doppler frequency, and 
	$ v_{rad,n}(t) $ denotes the additive white Gaussian noise.
	The existence of the jamming may result in a higher radar false alarm probability. Therefore, this paper aims at suppressing jamming and also ensuring the performance of sensing and communication via joint designing the ISAC waveform $ \mathbf{x}_{n} $ and the receive filter $ \mathbf{w}_{n}\in\mathbb{C}^{L} $.

	In Fig. \ref{fig:system_structure}, we depict the signal processing block diagram of the ISAC system. Firstly, based on the priori information on jamming and the communication symbols
	to be transmitted in the $ n $-th PRT, the ISAC system jointly designs the corresponding transmit signal $ \mathbf{x}_{n} $ and the receive filter $ \mathbf{w}_{n} $. Then, the communication user end acquires the modulated information symbols through the channel equalization and demodulation. During this period, the ISAC system receives the reflected target echo and jamming signal. Afterwards, the filter group $ \{\mathbf{w}_{n}e^{jn\theta_{1}},\cdots,\mathbf{w}_{n}e^{jn\theta_{M}}\},\theta_{i}=-\pi+2\pi (i-1)/(M-1), i=1,2,\cdots,M $ is applied to the radar receive signals to detect the target. In subsections II.A, II.B and II.C, the models of the radar, jamming and communication are described in detail.
	
	\subsection{Radar Model}
	It is widely known that the sensing performance of pulse radar depends on the coherent accumulation sum of each pulse of a CPI after pulse compression. This paper considers the joint design of the transmitted waveform and the receive filter in ISAC. Therefore, the sum of coherent accumulation of the ISAC pulse signal $ \{\mathbf{x}_{n}\}_{n=0}^{N-1} $ and the receive filter $ \{\mathbf{w}_{n}\}_{n=0}^{N-1} $ in a CPI depends on the sum of their aperiodic cross-correlation function.
	In the $n$-th PRT, the aperiodic cross-correlation function of the transmitted waveform $\mathbf{x}_{n}$ and the receive filter $\mathbf{w}_{n}$ can be defined as \cite{J. H. Wang-2021}
	\begin{equation}\label{key}
		C_{\mathbf{x}_{n},\mathbf{w}_{n}}(k)=\begin{cases}
			\sum\limits_{l=0}\limits^{L-k-1}x_{n,l}w_{n,l+k}^{*},\;0\leq k\leq L-1,\\
			\sum\limits_{l=0}\limits^{L+k-1}x_{n,l-k}w_{n,l}^{*},\;1-L\leq k\leq -1.
		\end{cases}
	\end{equation}
	To assure the sensing performance, low sidelobe levels should be achieved, which can be converted to minimize the ISL of $\mathbf{x}_{n}$ and $\mathbf{w}_{n}$ given by
	\begin{equation}\label{key}
		\text{ISL}(\mathbf{x}_{n},\mathbf{w}_{n})=\sum_{k\in\bm{\Omega} }\left|C_{\mathbf{x}_{n},\mathbf{w}_{n}}(k)\right|^{2},
	\end{equation}
	where $ \bm{\Omega}=\{1-L,\cdots,-1,1,\cdots,L-1\} $. Furthermore, (4) can be rewritten as
	\begin{equation}\label{key}
		\text{ISL}(\mathbf{x}_{n},\mathbf{w}_{n})=\sum_{k\in\bm{\Omega}}\left|\mathbf{x}_{n}^{H}\mathbf{U}_{k}\mathbf{w}_{n}\right|^{2},
	\end{equation}
	where $\mathbf{U}_{k},k\in\bm{\Omega}$, are Toeplitz matrices with the $ k $-th diagonal elements being $ 1 $ and $ 0 $ elsewhere \cite{J. Song-2016}.\par
	
	Then, the sensing performance can be improved by minimizing $ \text{ISL}(\mathbf{x}_{n},\mathbf{w}_{n}) $. In the following, we propose the SDMJ principle and establish the jamming model.
	
	\subsection{Jamming Model}
\begin{figure}[htbp]
		\centering
		\subfigure[]
		{\begin{minipage}{8cm}
				\centering
				\includegraphics[width=1\linewidth]{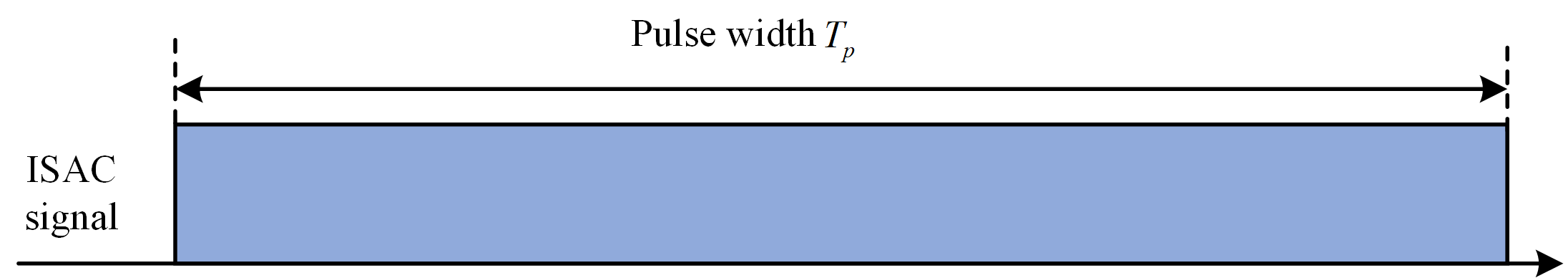}
		\end{minipage}}
		\subfigure[]
		{\begin{minipage}{8cm}
				\centering
				\includegraphics[width=1\linewidth]{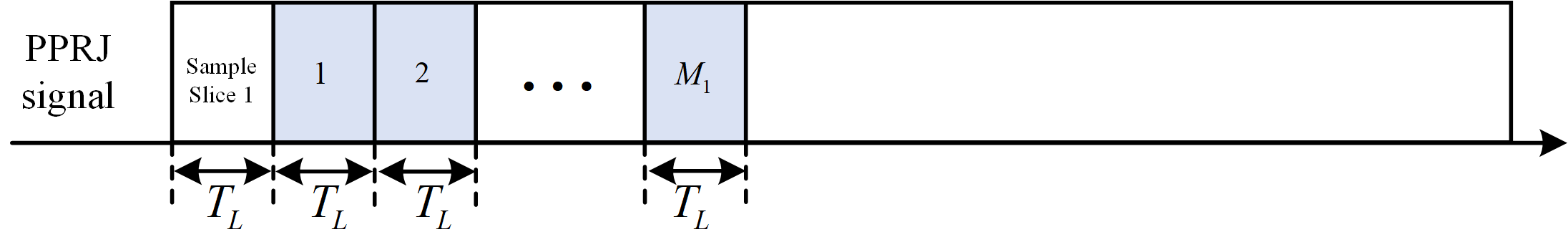}
		\end{minipage}}
		\subfigure[]
		{\begin{minipage}{8cm}
				\centering
				\includegraphics[width=1\linewidth]{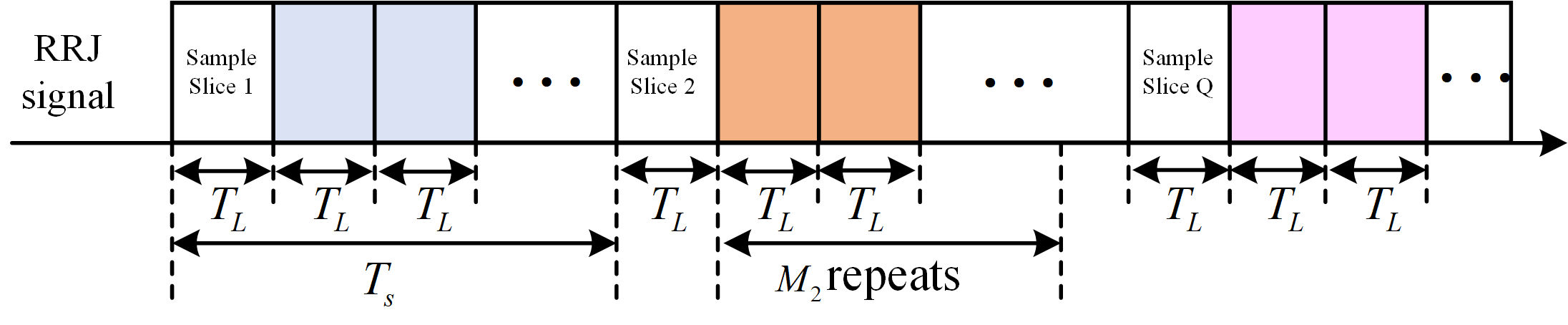}
		\end{minipage}}
		\caption{The principle of the two types of the SDMJ. (a) ISAC signal; (b) PPRJ signal; (c) RRJ signal.}	
		\label{fig:interference_principle}
	\end{figure}

	We mainly consider the problem of SDMJ suppression in this paper. Specifically, two types of the SDMJ are considered, i.e., partial pulse repeater jamming (PPRJ) and repetitive repeater jamming (RRJ) \cite{M. Ge-2021}. The principle of these two types of jamming is shown in Fig. \ref{fig:interference_principle}, where $ T_{p}=LT_{c} $ is assumed to be the pulse width of the ISAC signal.\par
	
	\par

	\begin{itemize}
		\item \textbf{PPRJ:} The DRFM jammer immediately intercepts, copies and forwards a part of the ISAC pulse signal to generate the PPRJ signal, as shown in Fig. \ref{fig:interference_principle}(b). $ T_{L} $ denotes the sampling time of the DRFM jammer, and the PPRJ signal is forwarded by $ M_{1}$ times.
		\item \textbf{RRJ:} The DRFM jammer firstly intercepts and samples the ISAC signal at a specific period. Then, it immediately forwards the sampling fragment until the next sampling time arrives. The operation will repeat until the pulse ends, as shown in Fig. \ref{fig:interference_principle}(c), where $T_{s} $ and  $Q=\lfloor T_{p}/T_{s}\rfloor $ represent the sampling interval and the sampling times, respectively, $M_{2}=\lfloor T_{s}/T_{L}\rfloor-1$ is the number of repeat.
	\end{itemize}

	It is assumed that the jammer keeps its characteristics constant within a CPI. According to the aforementioned principles of jamming, we can establish discrete jamming model as
	\begin{equation}\label{key}
		\mathbf{x}_{n,J}=\mathbf{J}\mathbf{x}_{n},\mathbf{J}\neq \mathbf{I}_{L},
	\end{equation}
	where $ \mathbf{x}_{n,J} $ represents the jamming, $\mathbf{J}$ denotes the jamming transfer matrix with size $ L\times L $. For the PPRJ signal, the transfer matrix $\mathbf{J}_{\text{PPRJ}}$ can be expressed as
	
	\begin{equation}\label{key}
		\mathbf{J}_{\text{PPRJ}}=\begin{bmatrix}
			M_{1}\begin{cases}
				\mathbf{I}_{c_{1}}\\
				\vdots\\
				\mathbf{I}_{c_{1}}
			\end{cases}&\begin{matrix}
				\mathbf{O}&\dots&\mathbf{O}\\
				\vdots&\ddots&\mathbf{O}\\
				\mathbf{O}&\dots&\mathbf{O}
			\end{matrix}\\
			\mathbf{O}&\begin{matrix}
				\mathbf{O}&\dots&\mathbf{O}
			\end{matrix}
		\end{bmatrix},
	\end{equation}
	where $c_{1}=\lfloor T_{L}/t_{s}\rfloor $ denotes sampling sequence length of jamming, $t_{s}$ represents the sampling time interval with the signal bandwidth as the sampling rate. Besides, the transfer matrix $\mathbf{J}$ of the RRJ signal can be expressed as
	\begin{equation}\label{key}
		\mathbf{J}_{\text{RRJ}}=\begin{bmatrix}
			\mathbf{D}&\mathbf{O}&\dots&\mathbf{O}\\
			\mathbf{O}&\mathbf{D}&\dots&\mathbf{O}\\
			\vdots&\vdots&\ddots&\mathbf{O}\\
			\mathbf{O}&\mathbf{O}&\dots&\mathbf{D}
		\end{bmatrix},	
		\mathbf{D}=\begin{bmatrix}
			M_{2}\begin{cases}
				\mathbf{I}_{c_{2}}\\
				\vdots\\
				\mathbf{I}_{c_{2}}
			\end{cases}&\begin{matrix}
				\dots&\mathbf{O}\\
				\ddots&\mathbf{O}\\
				\dots&\mathbf{O}
			\end{matrix}\\
			\mathbf{O}&\begin{matrix}
				\dots&\mathbf{O}
			\end{matrix}
		\end{bmatrix}
	\end{equation}
	where $ \mathbf{D} $ a matrice with size $L_{s}\times L_{s}$, $ L_{s}=\lfloor T_{s}/t_{s}\rfloor $, and $c_{2}=\lfloor T_{L}/t_{s}\rfloor $.\par
	
	Similarly, the IL of the jamming
	$ \mathbf{x}_{n,J} $ and the receive filter $ \mathbf{w}_{n} $ can be defined as
	\begin{equation}\label{key}
		\text{IL}(\mathbf{x}_{n,J},\mathbf{w}_{n})=\sum_{k\in\bm{\Omega}_{J}}\left|\mathbf{x}_{n,J}^{H}\mathbf{U}_{k}\mathbf{w}_{n}\right|^{2},
	\end{equation}
	where $ \bm{\Omega}_{J}=\{1-L,\cdots,-1,0,1,\cdots,L-1\} $.
	It is noted that $\mathbf{J}$ is often assumed to be known based on the cognitive method \cite{M. Ge-2021}, \cite{C. Zhou-2018}. As a result, the jamming can be suppressed by minimizing $ \text{IL}(\mathbf{x}_{n,J},\mathbf{w}_{n}) $. \par

	\subsection{Communication Model}
	At the communication receiving end, a multi-path time-invariant wireless channel is considered.
	For the $ n $-th PRT, the received single carrier signal after passing a channel is 
	\begin{equation}\label{key}
		y_{com,n}(t)=\sum_{i=1}^{P}h_{i}x_{n}(t-\tau_{i})+v_{com,n}(t),
	\end{equation}
	where  $ h_{i}\sim \mathcal{CN}(0,1),i=1,2,\cdots,P $ denotes the communication channel response of the $ i $-th path, $ P $ is the number of resolvable paths, $ \tau_{i} $ indicates delay of the $ i $-th path, and $ v_{com,n}(t) $ is the additive white Gaussian noise.
	Further, the discrete representation of the received signal at the communication user in $ L $ symbol times is given by
	\begin{equation}\label{key}
		\mathbf{y}_{com,n}=\mathbf{H}\mathbf{x}_{n}+\mathbf{v}_{com,n},
	\end{equation}
	where $ \mathbf{H}\in \mathbb{C}^{L\times L} $ is the effective channel matrix, which is assumed to be perfectly known and remains constant in a CPI, i.e.,
	\begin{equation}\label{key}
		\mathbf{H}=\sum_{i=1}^{P}h_{i}\bm{\Psi}_{l_{\tau_{i}}},\bm{\Psi}_{l_{\tau_{i}}}=\begin{bmatrix}
			\mathbf{O}_{l_{\tau_{i}}\times(L-l_{\tau_{i}})}&\mathbf{I}_{l_{\tau_{i}}}\\
			\mathbf{I}_{L-l_{\tau_{i}}}&\mathbf{O}_{(L-l_{\tau_{i}})\times l_{\tau_{i}}}
		\end{bmatrix},
	\end{equation}
	where $ l_{\tau_{i}}=\lfloor \tau_{i}/T_{c}\rfloor,i=1,2,\cdots,P $ is the number of the delay bins of the $ i $-th path.\par
	Furthermore, denote by $\mathbf{s}_{n}$ the desired communication waveform, where $\mathbf{s}_{n}=[s_{n,0},s_{n,1},\cdots,s_{n,L-1}]^{T} $, $s_{n,l}\in\mathcal{O}$, $ n=0,1,\cdots,N-1 $, $ l=0,1,\cdots,L-1 $, and $ \mathcal{O}$ denotes the set of the employed constellation points. For the communication purpose, we aim at designing a transmitted signal $\mathbf{x}_{n}$ which is very close to $\mathbf{s}_{n}$. For this, we introduce the following similarity constraint 
	\begin{equation}\label{key}
		\parallel \mathbf{x}_{n}-\mathbf{s}_{n}\parallel_{2}^{2}\leq\zeta,
	\end{equation}
	where $ \zeta $ is a parameter for controlling the degree of the similarity. Clearly, the closer $\mathbf{x}_{n}$ to $\mathbf{s}_{n}$, the better for its desirable properties.
	However, such a strict constraint would have severe impact on the performance of the sensing system, as well. Since the similarity constraint is related to the communication system only, a
	more flexible approach can be done by relaxing this constraint and letting it as a penalty term in the objective function of the optimization problem \cite{C. G. Tsinos-2021}. Therefore, the goal of this paper is to solve the following minimization problem
	\begin{equation}\label{key}
		\min_{\mathbf{x}_{n},\mathbf{w}_{n}}\;\rho\text{ISL}(\mathbf{x}_{n},\mathbf{w}_{n})+(1-\rho)\text{IL}(\mathbf{x}_{n,J},\mathbf{w}_{n})+\epsilon\parallel \mathbf{x}_{n}-\mathbf{s}_{n}\parallel_{2}^{2},
	\end{equation}
	where $\rho\in[0,1] $ is the weight factor that determines the weight between the sensing and jamming suppression performance in the ISAC system, $ \epsilon $ is the penalty parameter whose purpose is to control the degree of similarity between the designed ISAC waveform and the desired communication waveform.

	\section{Joint Design of the ISAC Waveform and Filter}
	In this section, we first construct the optimization model for a joint design of ISAC waveform and  filter. Then, we put forward the DMM algorithm to solve the formulated problem. Through the convergence and complexity analysis, the feasibility of the proposed algorithm is proved theoretically.
	
	\subsection{Problem Formulation}
	
	To achieve sensing and communication functions in the jamming environment,
	we establish the following optimization problem aiming to minimize $\text{ISL}(\mathbf{x}_{n},\mathbf{w}_{n}) $ and $ \text{IL}(\mathbf{x}_{n,J},\mathbf{w}_{n})$ and subject to the constraints of the similarity, energy, pulse compression peak level, and PAPR, i.e., 
	
	\begin{equation}\label{key}
		\mathcal{P}_{0}\begin{cases}
			\begin{aligned}
				\min_{\mathbf{x}_{n},\mathbf{w}_{n}}\;&\rho\text{ISL}(\mathbf{x}_{n},\mathbf{w}_{n})+(1-\rho)\text{IL}(\mathbf{x}_{n,J},\mathbf{w}_{n})&\\
				&+\epsilon\parallel \mathbf{x}_{n}-\mathbf{s}_{n}\parallel_{2}^{2}&\\
				\mbox{s.t.}\;\;\;&C_{1}:\parallel\mathbf{x}_{n}\parallel_{2}^{2}=L,\parallel\mathbf{w}_{n}\parallel_{2}^{2}=L\\
				& C_{2}:\mathbf{x}_{n}^{H}\mathbf{w}_{n}=a_{\max}&\\
				& C_{3}:\mathbf{x}_{n,J}^{H}\mathbf{w}_{n}=a_{\min}&\\
				& C_{4}:\text{PAPR}(\mathbf{x}_{n})\leq\gamma^{2},
			\end{aligned}
		\end{cases}
	\end{equation}
	where $C_{1} $ is the energy constraint of $ \mathbf{x}_{n} $ and $ \mathbf{w}_{n} $, which should be
	constrained to a given power $ L $. The constraint $C_{2}$ should be formulated to control
	the LPG, which is caused by the mismatched filter. Besides, to suppress jamming, the minimum peak constraint $C_{3}$ is introduced to limit the pulse compression peak of the jamming and the receive filter. $ C_{4} $ corresponds to the PAPR constraint. In practical systems, in order for the radio frequency amplifier to
	operate at its maximum efficiency and avoid nonlinear effect, waveforms with low PAPR are desirable. Therefore, $ C_{4} $ is introduced to limit the peak power of the ISAC waveform, where $ \gamma^{2} \geq 1$ is the upper bound of $ \text{PAPR}(\mathbf{x}_{n}) $, and the PAPR of $  \mathbf{x}_{n} $ is defined as  \cite{F. L. Wang-2021}
	\begin{equation}\label{key}
		\text{PAPR}(\mathbf{x}_{n})=\dfrac{\max\limits_{l=0,1,\cdots,L-1}\{|x_{n,l}|^{2}\}}{\frac{1}{L}\parallel \mathbf{x}_{n} \parallel_{2}^{2}}.
	\end{equation}\par
	For the special case of $ \gamma = 1 $, the PAPR constraint $ C_{4} $ becomes the constant-modulus (CM)	constraint, i.e.,
	\begin{equation}\label{key}
		|x_{n,l}|=1,l=0,1,\cdots,L-1.
	\end{equation}
	It is interesting to note that when the similarity constraint is not considered, (15) degenerates into a waveform design problem related to radar only, and the sensing and anti-jamming integrated waveform can be designed. Similarly, we can set $ \rho = 1 $ and ignore constraint $ C_{3} $ to
	realize the ISAC waveform design without jamming.

	\subsection{Proposed DMM Algorithm}
	Since  $\mathcal{P}_{0}$ is a non-convex problem, which is difficult to solve directly, we develop a DMM algorithm to solve the problem $\mathcal{P}_{0}$. To simplify the derivation, the two optimization vectors $ \mathbf{x}_{n}$ and $\mathbf{w}_{n} $ are reconstructed as one vector, i.e., $ \mathbf{z}_{n}=[\mathbf{x}_{n}^{T},\mathbf{w}_{n}^{T}]^{T} $. Similarly, let $ \mathbf{z}_{n,J}=[\mathbf{x}_{n,J}^{T},\mathbf{w}_{n}^{T}]^{T} $. Moreover, in order that the objective functions of the problem $\mathcal{P}_{0}$ is a quartic with respect to the variable $ \mathbf{z}_{n} $, we first disregard the similarity constraint.
	Then, the problem is derived under the framework of the majorization minimization (MM) \cite{J. Song-2016}. The problem $\mathcal{P}_{0}$ can be rewritten as
	\begin{equation}\label{key}
		\mathcal{P}_{1}\begin{cases}
			\begin{aligned}
				\min_{\mathbf{z}_{n}}\;&\rho\sum_{k\in\bm{\Omega}}|\mathbf{z}_{n}^{H}\widetilde{\mathbf{U}}_{k}\mathbf{z}_{n}|^{2}
				+(1-\rho)\sum_{k\in\bm{\Omega}_{J}}|\mathbf{z}_{n,J}^{H}\widetilde{\mathbf{U}}_{k}\mathbf{z}_{n}|^{2}&\\
				\mbox{s.t.}\;\;&\mathbf{z}_{n}=[\mathbf{x}_{n}^{T},\mathbf{w}_{n}^{T}]^{T},\;\mathbf{z}_{n,J}=[\mathbf{x}_{n,J}^{T},\mathbf{w}_{n}^{T}]^{T}&\\
				& C_{1}:\parallel\mathbf{x}_{n}\parallel_{2}^{2}=L,\parallel\mathbf{w}_{n}\parallel_{2}^{2}=L&\\
				& C_{2}:\mathbf{z}_{n}^{H}\mathbf{\Gamma}\mathbf{z}_{n}=2a_{\max}&\\
				& C_{3}:(\bm{\mathcal{J}}\mathbf{z}_{n})^{H}\mathbf{\Gamma}\bm{\mathcal{J}}\mathbf{z}_{n}=2a_{\min}&\\
				& C_{4}:\text{PAPR}(\mathbf{x}_{n})\leq\gamma^{2},				
			\end{aligned}
		\end{cases}
	\end{equation}
	where
	\begin{equation}\label{key}
		\widetilde{\mathbf{U}}_{k}=\begin{bmatrix}
			\mathbf{O}&\mathbf{U}_{k}\\
			\mathbf{O}&\mathbf{O}
		\end{bmatrix},\;\;
		\mathbf{\Gamma}=\begin{bmatrix}
			\mathbf{O}&\mathbf{I}\\
			\mathbf{I}&\mathbf{O}
		\end{bmatrix},\;\;\bm{\mathcal{J}}=\begin{bmatrix}
			\mathbf{J}&\mathbf{O}\\
			\mathbf{O}&\mathbf{I}
		\end{bmatrix}.
	\end{equation}\par
	Since $ \mathbf{z}_{n}^{H}\widetilde{\mathbf{U}}_{k}\mathbf{z}_{n}=\mathbf{Tr}(\widetilde{\mathbf{U}}_{k}\mathbf{z}_{n}\mathbf{z}_{n}^{H}) $, the minimization problem of $\mathcal{P}_{1} $ can be simplified as
	\begin{equation}\label{key}
		\min_{\mathbf{Z}_{n}}\text{vec}(\mathbf{Z}_{n})^{H}\mathbf{W}\text{vec}(\mathbf{Z}_{n}),
	\end{equation}
	where $\mathbf{Z}_{n}=\mathbf{z}_{n}\mathbf{z}_{n}^{H}$, $\mathbf{W}= \rho\mathbf{A}+(1-\rho)\mathbf{B} $. Herein, 
	\begin{equation}\label{key}
		\mathbf{A}=\sum_{k\in\bm{\Omega}}\text{vec}(\widetilde{\mathbf{U}}_{k})\text{vec}(\widetilde{\mathbf{U}}_{k})^{H},
	\end{equation}
	\begin{equation}\label{key}
		\mathbf{B}=\sum_{k\in\bm{\Omega}_{J}}\text{vec}(\widetilde{\mathbf{U}}_{J,k})\text{vec}(\widetilde{\mathbf{U}}_{J,k})^{H},
	\end{equation}
	and
	\begin{equation}\label{key}
		\widetilde{\mathbf{U}}_{J,k}=\begin{bmatrix}
			\mathbf{O}&\mathbf{J}^{H}\mathbf{U}_{k}\\
			\mathbf{O}&\mathbf{O}
		\end{bmatrix}.
	\end{equation}\par
	Please see  Appendix A for the proof. $ \hfill\blacksquare $\par
	Obviously, (20) is a quadratic function corresponding to $\mathbf{Z}_{n}$, and $\mathbf {W}$ is a Hermitian matrix. To solve the problem (20) using the MM method, the
	key step is to find a majorization function of the objective function. For that purpose, we need the following result. 
	\begin{Lemma}\cite{J. Song-2015}
		Let $\mathbf{L}$ be an $n \times n$ Hermitian matrix and $\mathbf{M}$ be another $n\times n$ Hermitian matrix such that $\mathbf{M}\succeq\mathbf{L}$. Then for any point $\mathbf{x}_{0}\in\mathbb{C}^{n}$, the quadratic function $\mathbf{x}^{H}\mathbf{L}\mathbf{x}$ is majorized by $ \mathbf{x}^{H}\mathbf{M}\mathbf{x}+2\text{Re}\{\mathbf{x}^{H}(\mathbf{L}-\mathbf{M})\mathbf{x}_{0}\}+\mathbf{x}^{H}_{0}(\mathbf{M}-\mathbf{L})\mathbf{x}_{0} $ at $\mathbf{x}_{0}$.
	\end{Lemma}\par
	
	Then, we can construct an optimization function for (20) by selecting the matrix $\mathbf{M}=\lambda_{u}\mathbf{I}$, where $ \lambda_{u}=\lambda_{\max}(\mathbf{W})=\rho\lambda_{\max}(\mathbf{A})+(1-\rho)\lambda_{\max}(\mathbf{B})$ is the maximum eigenvalue of the matrix $\mathbf{W}$. Owing to the special structures of $ \mathbf{A} $ and $ \mathbf{B} $, it can be shown that $ \lambda_{\max}(\mathbf{A}) $ and $ \lambda_{\max}(\mathbf{B}) $ can be computed efficiently in closed form.
	
	\begin{Lemma}
		Let $ \mathbf{A} $ and $ \mathbf{B} $ be two matrices defined in (21) and (22), respectively. Then, the maximum eigenvalues of $ \mathbf{A} $ and $ \mathbf{B} $ are given by $ \lambda_{\max}(\mathbf{A})=L-1 $, and $ \lambda_{\max}(\mathbf{B})=\max\limits_{k}\{\text{vec}(\mathbf{J}^{H}\mathbf{U}_{k})^{H}\text{vec}(\mathbf{J}^{H}\mathbf{U}_{k})|k\in\bm{\Omega}_{J}\} $.
	\end{Lemma} 
	Please see  Appendix B for the proof. $ \hfill\blacksquare $\par
	
	Thus, the maximum eigenvalue of $ \mathbf{W} $ is given by
	\begin{equation}\label{key}
		\begin{split}
			\lambda_{u}=&\rho(L-1)\\
			&+\max_{k\in\bm{\Omega}_{J}}\{(1-\rho)\text{vec}(\mathbf{J}^{H}\mathbf{U}_{k})^{H}\text{vec}(\mathbf{J}^{H}\mathbf{U}_{k})\}.
		\end{split}
	\end{equation}
	Let $ \mathbf{Z}_{n}^{(t)}=\mathbf{z}_{n}^{(t)}(\mathbf{z}_{n}^{(t)})^{H} $.
	Then, the objective function in (18) can be majorized by the following function at $\mathbf{Z}_{n}^{(t)} $
	\begin{equation}\label{key}
		\begin{split}
			u_{1}(\mathbf{Z}_{n},\mathbf{Z}_{n}^{(t)})
			=&\lambda_{u}\text{vec}(\mathbf{Z}_{n})^{H}\text{vec}(\mathbf{Z}_{n})\\
			&+2\text{Re}\left\{\text{vec}(\mathbf{Z}_{n})^{H}(\mathbf{W}-\lambda_{u}\mathbf{I})\text{vec}(\mathbf{Z}_{n}^{(t)})\right\}\\
			&+\text{vec}(\mathbf{Z}_{n}^{(t)})^{H}(\lambda_{u}\mathbf{I}-\mathbf{W})\text{vec}(\mathbf{Z}_{n}^{(t)}).
		\end{split}
	\end{equation}
	Since $\text{vec}(\mathbf{Z}_{n})^{H}\text{vec}(\mathbf{Z}_{n})=(\mathbf{z}_{n}^{H}\mathbf{z}_{n})^{2}=4L^{2} $, the first term of (25) is a constant. Thus, (25) can be further simplified to
	\begin{equation}\label{key}
		\begin{split}
			&u_{1}(\mathbf{Z}_{n},\mathbf{Z}_{n}^{(t)})\\
			=&2\text{Re}\left\{\text{vec}(\mathbf{Z}_{n})^{H}(\mathbf{W}-\lambda_{u}\mathbf{I})\text{vec}(\mathbf{Z}_{n}^{(t)})\right\}+8\lambda_{u}L^{2}\\
			&-\rho\sum_{k\in\bm{\Omega}}|C_{\mathbf{x}_{n},\mathbf{w}_{n}}^{(t)}(k)|^{2}-(1-\rho)\sum_{k\in\bm{\Omega}_{J}}|C_{\mathbf{x}_{n,J},\mathbf{w}_{n}}^{(t)}(k)|^{2},
		\end{split}
	\end{equation}
	where $C_{\mathbf{x}_{n},\mathbf{w}_{n}}^{(t)}(k) $ represents the aperiodic cross-correlation of $\mathbf{x}_{n}$ and $\mathbf{w}_{n}$ at iteration $ t $, and $ C_{\mathbf{x}_{n,J},\mathbf{w}_{n}}^{(t)}(k) $ denotes the aperiodic cross-correlation of $ \mathbf{x}_{n,J} $ and $\mathbf{w}_{n}$ at iteration $ t $. After ignoring the constant terms, the majorized problem of (25) is given by
	\begin{equation}\label{key}
		\min_{\mathbf{Z}_{n}}\text{Re}\left\{\text{vec}(\mathbf{Z}_{n})^{H}(\mathbf{W}-\lambda_{u}\mathbf{I})\text{vec}(\mathbf{Z}_{n}^{(t)})\right\}.
	\end{equation}
	Then, substituting $ \mathbf{W} $ into (27), the majorized problem of (27) is further transformed into
	\begin{equation}\label{key}
		\min_{\mathbf{z}_{n}}\text{Re}\{\mathbf{z}_{n}^{H}(\mathbf{Q}-\lambda_{u}\mathbf{z}_{n}^{(t)}(\mathbf{z}_{n}^{(t)})^{H})\mathbf{z}_{n}\},
	\end{equation}
	where 
	\begin{equation}\label{key}
		\mathbf{Q}
		=\begin{bmatrix}
			\mathbf{O}&\rho\bm{\Phi}\\
			\mathbf{O}&\mathbf{O}
		\end{bmatrix}
		+\begin{bmatrix}
			\mathbf{O}&(1-\rho)\mathbf{J}^{H}\bm{\Phi}_{J}\\
			\mathbf{O}&\mathbf{O}
		\end{bmatrix},
	\end{equation}
	and
	\begin{equation}\label{key}
		\bm{\Phi}=\sum\limits_{k\in\bm{\Omega}}C_{\mathbf{w}_{n},\mathbf{x}_{n}}^{(t)}(-k)\mathbf{U}_{k},\;\bm{\Phi}_{J}=\sum\limits_{k\in\bm{\Omega}_{J}}C_{\mathbf{w}_{n},\mathbf{x}_{n,J}}^{(t)}(-k)\mathbf{U}_{k}.
	\end{equation}

	Please see  Appendix C for the proof. $ \hfill\blacksquare $\par
	We can see that the $ \bm{\Phi} $ and $ \bm{\Phi}_{J} $ are Hermitian Toeplitz matrices, thus we can introduce FFT and IFFT to efficiently compute them by the \emph{Lemma} 4 in \cite{J. Song-2016}, we easily have
	\begin{equation}\label{key}
		\begin{split}
			\mathbf{Q}
			=&\begin{bmatrix}
				\mathbf{O}&\dfrac{\rho}{2L}\mathbf{F}_{:,1:L}^{H}\text{Diag}(\bm{\mu})\mathbf{F}_{:,1:L}\\
				\mathbf{O}&\mathbf{O}
			\end{bmatrix}\\
			&+\begin{bmatrix}
				\mathbf{O}&\dfrac{1-\rho}{2L}\mathbf{J}^{H}\mathbf{F}_{:,1:L}^{H}\text{Diag}(\bm{\mu}_{J})\mathbf{F}_{:,1:L}\\
				\mathbf{O}&\mathbf{O}
			\end{bmatrix},
		\end{split}
	\end{equation}
	where $ \bm{\mu}=\mathbf{F}(\bm{\omega}\circ\mathbf{c}) ,\bm{\mu}_{J}=\mathbf{F}(\bm{\omega}_{J}\circ\mathbf{c}_{J})$, $ \bm{\omega}=[0,\bm{1}_{ L-1},0,\bm{1}_{ L-1}]^{T} $,$ \bm{\omega}_{J}=[1,\bm{1}_{ L-1},0,\bm{1}_{ L-1}]^{T} $, $ \bm{1}_{L-1}$ is an all ones row vector of length $ L-1 $, and
	\begin{equation}\label{key}
		\mathbf{c}=\mathbf{F}^{H}\left((\mathbf{F}[(\mathbf{w}_{n}^{(t)})^{T},\bm{0}_{1\times L}]^{T})^{*}\circ(\mathbf{F}[(\mathbf{x}_{n}^{(t)})^{T},\bm{0}_{1\times L}]^{T})\right),
	\end{equation}
	\begin{equation}\label{key}
		\mathbf{c}_{J}=\mathbf{F}^{H}\left((\mathbf{F}[(\mathbf{w}_{n}^{(t)})^{T},\bm{0}_{1\times L}]^{T})^{*}\circ(\mathbf{F}[(\mathbf{x}_{n,J}^{(t)})^{T},\bm{0}_{1\times L}]^{T})\right).
	\end{equation}\par

	Since $\mathbf{Q}-\lambda_{u}\mathbf{z}_{n}^{(t)}(\mathbf{z}_{n}^{(t)})^{H}  $ in (28) is not a Hermitian matrix, the optimization
	problem (28) is not a traditional unimodular quadratic programming (UQP) defined in \cite{M. Soltanalian-2014}. To solve this problem, we can add the conjugate term $ \mathbf{Q}^{H}-\lambda_{u}\mathbf{z}_{n}^{(t)}(\mathbf{z}_{n}^{(t)})^{H} $ to (28), which does not change the optimization results. Then, (28) can be equivalently written as 
	\begin{equation}\label{key}
		\min_{\mathbf{z}_{n}}\text{Re}\{\mathbf{z}_{n}^{H}(\mathbf{Q}+\mathbf{Q}^{H}-2\lambda_{u}\mathbf{z}_{n}^{(t)}(\mathbf{z}_{n}^{(t)})^{H})\mathbf{z}_{n}\}.
	\end{equation}\par
	To effectively control the LPG and suppress jamming, we consider the constraints $C_{2}$ and $C_{3} $. Without loss of generality, the constraints $C_{2}$ and $C_{3} $ can be relaxed by regarding them as penalty terms in the objective function  \cite{Y. Chen-2024}, i.e.,
	\begin{equation}\label{key}
		g(\mathbf{z}_{n})=|\mathbf{z}_{n}^{H}\mathbf{\Gamma}\mathbf{z}_{n}-2a_{\max}|^{2}+|(\bm{\mathcal{J}}\mathbf{z}_{n})^{H}\mathbf{\Gamma}\bm{\mathcal{J}}\mathbf{z}_{n}-2a_{\min}|^{2}.
	\end{equation}
	Then, the objective function can be further given as
	\begin{equation}\label{key}
		\begin{aligned}
			&\text{Re}\{\mathbf{z}_{n}^{H}(\mathbf{Q}+\mathbf{Q}^{H}-2\lambda_{u}\mathbf{z}_{n}^{(t)}(\mathbf{z}_{n}^{(t)})^{H}\\
			&-2\beta_{1}a_{\max}\mathbf{\Gamma}^{H}-2\beta_{2}a_{\min}(\bm{\mathcal{J}}^{H}\mathbf{\Gamma}\bm{\mathcal{J}})^{H})\mathbf{z}_{n}\},&
		\end{aligned}
	\end{equation}
	where $ \beta_{1}$ and $\beta_{2} $ are the penalty factors with $ \beta_{1}+\beta_{2}=1 $, which control the weighs of LPG and the jamming peak.\par 
	Furthermore, we consider the similarity constraint. 
	Let $\tilde{\mathbf{s}}_{n}=[\mathbf{s}_{n}^{T},\mathbf{0}_{1\times L}]^{T} $. Then, the penalty term of the objective function in problem $ \mathcal{P}_{0} $ can be equivalently written as $ \parallel \mathbf{z}_{n}-\tilde{\mathbf{s}}_{n}\parallel_{2}^{2} $,
	which can be transformed into 
	\begin{equation}\label{key}
		-2\text{Re}\{\mathbf{z}_{n}^{H}(\tilde{\mathbf{s}}_{n}\tilde{\mathbf{s}}_{n}^{H})\mathbf{z}_{n}\}.
	\end{equation}
	Therefore, the  problem $ \mathcal{P}_{0} $ can be further written as
	\begin{equation}\label{key}
		\mathcal{P}_{2}\begin{cases}
			\begin{aligned}
				\min_{\mathbf{z}_{n}}\;&\text{Re}\{\mathbf{z}_{n}^{H}\mathbf{R}\mathbf{z}_{n}\}&\\
				\mbox{s.t.}\;\;&\mathbf{z}_{n}=[\mathbf{x}_{n}^{T},\mathbf{w}_{n}^{T}]^{T}&\\
				&C_{1}:\parallel\mathbf{x}_{n}\parallel_{2}^{2}=L,\parallel\mathbf{w}_{n}\parallel_{2}^{2}=L&\\
				&C_{4}:\text{PAPR}(\mathbf{x}_{n})\leq\gamma^{2},
			\end{aligned}
		\end{cases}
	\end{equation}
	where
	\begin{equation}\label{key}
		\begin{split}
			\mathbf{R}=&\mathbf{Q}+\mathbf{Q}^{H}-2\lambda_{u}\mathbf{z}_{n}^{(t)}(\mathbf{z}_{n}^{(t)})^{H}\\
			&-2\beta_{1}a_{\max}\mathbf{\Gamma}^{H}-2\beta_{2}a_{\min}(\bm{\mathcal{J}}^{H}\mathbf{\Gamma}\bm{\mathcal{J}})^{H}-\epsilon\tilde{\mathbf{s}}_{n}\tilde{\mathbf{s}}_{n}^{H}. 
		\end{split}
	\end{equation}
	Obviously, $  \mathbf{R}$ is a Hermitian matrix. Thus, the $ \text{Re}\{\cdot\}$ of the objective function of $ \mathcal{P}_{2}$ can be further removed. According to Lemma 1, we can choose 
	\begin{equation}\label{key}
		\mathbf{M}= \lambda_{v}\mathbf{I},
	\end{equation}
	where $ \lambda_{v}=\mathbf{Tr}(\mathbf{Q}+\mathbf{Q}^{H})=\mathbf{Tr}(\mathbf{Q})+\mathbf{Tr}(\mathbf{Q}^{H}) $ denotes the upper bound of the maximum eigenvalue $\lambda_{\max}(\mathbf{R})$ of $ \mathbf{R}$, i.e.,
	\begin{equation}\label{key}
		\lambda_{v}\geq\lambda_{\max}(\mathbf{Q}+\mathbf{Q}^{H})\geq\lambda_{\max}(\mathbf{R}).
	\end{equation}
	By observing (31), it can be found that $ \mathbf{Q} $ is an upper triangular matrix with zero principal diagonal elements, and thus $ \lambda_{v}=0 $.
	Then, the objective function in problem $\mathcal{P}_{2} $ can be optimized by the following function
	\begin{equation}\label{key}
		u_{2}(\mathbf{z}_{n},\mathbf{z}_{n}^{(t)})=
		2\text{Re}\{\mathbf{z}_{n}^{H}\mathbf{R}\mathbf{z}_{n}^{(t)}\}\\
		+(\mathbf{z}_{n}^{(t)})^{H}\mathbf{R}\mathbf{z}_{n}^{(t)}.
	\end{equation}\par
	Note that although we have applied the majorization-minimization scheme twice at the point $ \mathbf{z}_{n}^{(t)} $,  it can be viewed as directly majorizing the objective function of $ \mathcal{P}_{0} $ at $ \mathbf{z}_{n}^{(t)} $ by the following function
	\begin{equation}\label{key}
		\begin{split}
			&u(\mathbf{z}_{n},\mathbf{z}_{n}^{(t)})\\
			&=u_{2}(\mathbf{z}_{n},\mathbf{z}_{n}^{(t)})+8\lambda_{u}L^{2}\\
			&\;\;\;\;-\rho\sum_{k\in\bm{\Omega}}|C_{\mathbf{x}_{n},\mathbf{w}_{n}}^{(t)}(k)|^{2}-(1-\rho)\sum_{k\in\bm{\Omega}_{J}}|C_{\mathbf{x}_{n,J},\mathbf{w}_{n}}^{(t)}(k)|^{2}\\
			&=-2\text{Re}\{\mathbf{z}_{n}^{H}\mathbf{P}(\mathbf{z}_{n}^{(t)})\}+16\lambda_{u}L^{2}\\
			&\;\;\;\;-\rho\sum_{k\in\bm{\Omega}}\left(|C_{\mathbf{x}_{n},\mathbf{w}_{n}}^{(t)}(k)|^{2}+|C_{\mathbf{w}_{n},\mathbf{x}_{n}}^{(t)}(k)|^{2}\right)\\
			&\;\;\;\;-(1-\rho)\sum_{k\in\bm{\Omega}_{J}}\left(|C_{\mathbf{x}_{n,J},\mathbf{w}_{n}}^{(t)}(k)|^{2}+|C_{\mathbf{w}_{n},\mathbf{x}_{n,J}}^{(t)}(k)|^{2}\right)\\
			&\;\;\;\;-4\beta_{1}a_{\max}^{2}-4\beta_{2}a_{\min}^{2}-2L\epsilon\tilde{\mathbf{s}}_{n}\tilde{\mathbf{s}}_{n}^{H},\\
		\end{split}
	\end{equation}
	where
	\begin{equation}\label{key}
		\begin{split}
			\mathbf{P}(\mathbf{z}_{n}^{(t)})=&4\lambda_{u}L\mathbf{z}_{n}^{(t)}+\epsilon\tilde{\mathbf{s}}_{n}\tilde{\mathbf{s}}_{n}^{H}\mathbf{z}_{n}^{(t)}-(\mathbf{Q}+\mathbf{Q}^{H})\mathbf{z}_{n}^{(t)}\\
			&+(2\beta_{1}a_{\max}\mathbf{\Gamma}^{H}+2\beta_{2}a_{\min}(\bm{\mathcal{J}}^{H}\mathbf{\Gamma}\bm{\mathcal{J}})^{H})\mathbf{z}_{n}^{(t)},
		\end{split}
	\end{equation}
	and
	\begin{equation}\label{key}
		\begin{split}
			&(2\beta_{1}a_{\max}\mathbf{\Gamma}^{H}+2\beta_{2}a_{\min}(\bm{\mathcal{J}}^{H}\mathbf{\Gamma}\bm{\mathcal{J}})^{H}+\epsilon\tilde{\mathbf{s}}_{n}\tilde{\mathbf{s}}_{n}^{H})\mathbf{z}_{n}^{(t)}\\
			&=\begin{bmatrix}
				2\beta_{1}a_{\max}\mathbf{w}_{n}+2\beta_{2}a_{\min}\mathbf{J}^{H}\mathbf{w}_{n}+\epsilon\mathbf{s}_{n}\mathbf{s}_{n}^{H}\mathbf{x}_{n}\\
				2\beta_{1}a_{\max}\mathbf{x}_{n}+2\beta_{2}a_{\min}\mathbf{J}\mathbf{x}_{n}\end{bmatrix},
		\end{split}    	
	\end{equation}
	\begin{equation}\label{key}
		\begin{split}
			&(\mathbf{Q}+\mathbf{Q}^{H})\mathbf{z}_{n}^{(t)}\\
			&=\begin{bmatrix}
				\dfrac{\rho}{2L}\mathbf{F}_{:,1:L}^{H}(\bm{\mu}\circ\mathbf{f}_{\mathbf{w}_{n}})+\dfrac{1-\rho}{2L}\mathbf{J}^{H}\mathbf{F}_{:,1:L}^{H}(\bm{\mu}_{J}\circ\mathbf{f}_{\mathbf{w}_{n}})\\
				\dfrac{\rho}{2L}\mathbf{F}_{:,1:L}^{H}(\bm{\mu}^{*}\circ\mathbf{f}_{\mathbf{x}_{n}})+\dfrac{1-\rho}{2L}\mathbf{F}_{:,1:L}^{H}(\bm{\mu}_{J}^{*}\circ\mathbf{f}_{\mathbf{x}_{n,J}})
			\end{bmatrix},
		\end{split}
	\end{equation}
	where $\mathbf{f}_{\mathbf{x}_{n}}=\mathbf{F}[(\mathbf{x}_{n}^{(t)})^{T},\bm{0}_{1\times L}]^{T},\mathbf{f}_{\mathbf{w}_{n}}=\mathbf{F}[(\mathbf{w}_{n}^{(t)})^{T},\bm{0}_{1\times L}]^{T}$, $\mathbf{f}_{\mathbf{x}_{n,J}}=\mathbf{F}[(\mathbf{x}_{n,J}^{(t)})^{T},\bm{0}_{1\times L}]^{T}$.
	
	After neglecting the constant terms of $ u(\mathbf{z}_{n},\mathbf{z}_{n}^{(t)}) $, the objective function of the optimization problem $ \mathcal{P}_{0} $ is given by
	\begin{equation}\label{key}
		\min_{\mathbf{z}_{n}}\;-2\text{Re}\{\mathbf{z}_{n}^{H}\mathbf{P}(\mathbf{z}_{n}^{(t)})\}.
	\end{equation}
	Then, the optimization problem $ \mathcal{P}_{0} $ can be rewritten as
	\begin{equation}\label{key}
		\mathcal{P}_{3}\begin{cases}
			\begin{aligned}
				\min_{\mathbf{z}_{n}}\;&\parallel\mathbf{z}_{n}-\mathbf{P}(\mathbf{z}_{n}^{(t)})\parallel_{2}&\\
				\mbox{s.t.}\;\;&\mathbf{z}_{n}=[\mathbf{x}_{n}^{T},\mathbf{w}_{n}^{T}]^{T}&\\
				& C_{1}:\parallel\mathbf{x}_{n}\parallel_{2}^{2}=L,\parallel\mathbf{w}_{n}\parallel_{2}^{2}=L&\\
				& C_{4}:\text{PAPR}(\mathbf{x}_{n})\leq\gamma^{2},
			\end{aligned}
		\end{cases}
	\end{equation}
	Obviously, the objective function of $ \mathcal{P}_{3} $ is a linear optimization problem that yields a closed-form solution of $ \mathbf{z}_{n} $. When constraints are not considered, we can directly obtain the solution of the $ \mathbf{x}_{n} $
	\begin{equation}\label{key}
		x_{n,l}^{(t+1)}=|x_{n,l}^{(t+1)}|e^{j\text{arg}(\mathbf{P}_{\mathbf{x}}(\mathbf{z}_{n,l}^{(t)}))},l=0,1,\cdots,L-1,
	\end{equation}
	where $ \mathbf{P}_{\mathbf{x}}(\mathbf{z}_{n,l}^{(t)}) $ is the $ l $-th element of the $ \mathbf{P}_{\mathbf{x}}(\mathbf{z}_{n}^{(t)}) $, and $ \mathbf{P}_{\mathbf{x}}(\mathbf{z}_{n}^{(t)}) $ denotes the $ 1 $ to $ L $ rows of the $ \mathbf{P}(\mathbf{z}_{n}^{(t)}) $. 
	Meanwhile, by considering the energy constraint $ C_{1} $, we can obtain an optimal solution of $ \mathbf{w}_{n} $ as
	\begin{equation}\label{key}
		\mathbf{w}_{n}^{(t+1)}=\sqrt{L/\parallel \mathbf{P}_{\mathbf{w}}(\mathbf{z}_{n}^{(t)})\parallel_{2}^{2}}\mathbf{P}_{\mathbf{w}}(\mathbf{z}_{n}^{(t)}),
	\end{equation}
	where $ \mathbf{P}_{\mathbf{w}}(\mathbf{z}_{n}^{(t)}) $ indicates the $ L+1 $ to $ 2L $ rows of the $ \mathbf{P}(\mathbf{z}_{n}^{(t)}) $. \par
	
	Moreover, to address the PAPR problem of the ISAC waveform, we propose the modulus constraint to design the optimal ISAC waveform, which is written as 
	\begin{equation}\label{key}
		|x_{n,l}^{(t+1)}|=\min\{\delta|\mathbf{P}_{\mathbf{x}}(\mathbf{z}_{n,l}^{(t)})|,\gamma\}.
	\end{equation}
	In addition, to satisfy the energy constraint $ C_{1} $, we have
	\begin{equation}\label{key}
		\begin{split}
			&\delta\in\left\{\delta\mid f(\delta)=0,\delta\in\left(0,\delta_{u}\right)\right\},\\
			&f(\delta)= \sum_{l=0}^{L-1}\min\{\gamma^{2},\delta^{2}|\mathbf{P}_{\mathbf{x}}(\mathbf{z}_{n,l}^{(t)})|^{2}\}-L,
		\end{split}
	\end{equation}
	where $ \delta_{u}=\gamma/\min\{|\mathbf{P}_{\mathbf{x}}(\mathbf{z}_{n,l}^{(t)})|,|\mathbf{P}_{\mathbf{x}}(\mathbf{z}_{n,l}^{(t)})|\neq 0\} $.
	Since the function $ f(\delta) $ is monotonically increasing in the interval $(0,\delta_{u}) $  and $ f(0)< 0 $, $ \delta $ can be determined by the bisection method (BM) in Algorithm 1.
	\begin{algorithm}[h]
		\caption{BM for Solving $ \delta $ in (52)} %算法的名字
		\hspace*{0.02in} {\bf Input:} %算法的输入， \hspace*{0.02in}用来控制位置，同时利用 \\ 进行换行
		Search interval: $ (\delta_{1},\delta_{2}) $, set $ \text{eps}=1\times 10^{-12}, \delta_{1}=0, \delta_{2}=\delta_{u}$, and $ \delta\in(\delta_{1},\delta_{2}) $ ;\\
		\hspace*{0.02in} {\bf Output:} %算法的结果输出
		$ \delta $;
		\begin{algorithmic}[1]
			\While{$ |\delta_{1}-\delta_{2}|\geq\text{eps}$}
			\State$ \delta=(\delta_{1}+\delta_{2})/2 $;
			\If{$ f(\delta)>0 $} % If 语句，需要和EndIf对应
			\State $ \delta_{2}=\delta $;
			\Else 
			\State $ \delta_{1}=\delta $;
			\EndIf
			\State \textbf{end if}
			\EndWhile
			\State \textbf{end while}
		\end{algorithmic}
	\end{algorithm}\par
	
	Combining (49) and (51), the optimal solution $ \mathbf{x}_{n} $ of the problem $ \mathcal{P}_{0} $ can be obtained as
	\begin{equation}\label{key}
		\tilde{x}_{n,l}^{(t+1)}=\begin{cases}
			\delta|\mathbf{P}_{\mathbf{x}}(\mathbf{z}_{n,l}^{(t)})|e^{j\text{arg}(\mathbf{P}_{\mathbf{x}}(\mathbf{z}_{n,l}^{(t)}))},&\delta|\mathbf{P}_{\mathbf{x}}(\mathbf{z}_{n,l}^{(t)})|\in[0,\gamma),\\
			\gamma e^{j\text{arg}(\mathbf{P}_{\mathbf{x}}(\mathbf{z}_{n,l}^{(t)}))},&\delta|\mathbf{P}_{\mathbf{x}}(\mathbf{z}_{n,l}^{(t)})|\geq\gamma.
		\end{cases}
	\end{equation}
	In this way, we obtain the solution $ \mathbf{x}_{n}=\tilde{\mathbf{x}}_{n} $  for the optimization problem $ \mathcal{P}_{0} $.\par
	
	In summary, the joint design problem of the transmitted waveform and the receive filter is transformed into a single vector iterative optimization problem, and the MM algorithm is used in each iteration to simplify the problem. According to
	the principle of the MM algorithm \cite{J. Song-2016}, it is known that the DMM algorithm proposed in this paper is monotonic and convergent, and we will give the proof in the sequel. Algorithm 2 summarizes the detailed procedures for solving the problem $ \mathcal{P}_{0} $. The algorithm can be terminated if the relative change of the variables is smaller than a predefined threshold $ \eta $, i.e.,
	\begin{equation}\label{key}
		\text{eps}=\parallel\mathbf{x}_{n}^{(t+1)}-\mathbf{x}_{n}^{(t)}\parallel_{2}+\parallel\mathbf{w}_{n}^{(t+1)}-\mathbf{w}_{n}^{(t)}\parallel_{2}\leq\eta.
	\end{equation}\par
	\begin{algorithm}[h]
		\caption{DMM Algorithm for Solving $ \mathcal{P}_{0} $} %算法的名字
		\hspace*{0.02in} {\bf Initialize:} $ t=0,\mathbf{x}_{n}^{(0)},\mathbf{w}_{n}^{(0)},\mathbf{s}_{n}$, and $ \mathbf{x}_{n}^{(0)}= \mathbf{s}_{n} $;\\
		\hspace*{0.02in} {\bf Input:} %算法的输入， \hspace*{0.02in}用来控制位置，同时利用 \\ 进行换行
		$\gamma,\rho,\beta_{1},\beta_{2},\epsilon$;\\
		\hspace*{0.02in} {\bf Output:} %算法的结果输出
		$ \mathbf{x}_{n}^{(t+1)} ,\mathbf{w}_{n}^{(t+1)} $;
		\begin{algorithmic}[1]
			\While{$\text{eps}> \eta $}
			\State Compute $ \mathbf{w}_{n}^{(t+1)} $ according to equation (50).
			\State Calculate $ \delta $ by using the Algorithm 1, and compute $ \mathbf{x}_{n}^{(t+1)} $ according to equation (53).
			\State Compute $ \text{eps} $ according to equation (54).
			\State $ t\gets t+1 $.
			\EndWhile
			\State \textbf{end while}
		\end{algorithmic}
	\end{algorithm}
	Since the convergence rate of the MM algorithm is usually related to the constructed optimization functions, an acceleration method was proposed in \cite{J. Song-2015} based on the squared iterative method (SQUAREM), to guarantee the convergence of the original algorithm. To further improve the convergence rate of the DMM algorithm, the SQUAREM can also be used in this paper to accelerate the DMM algorithm.\par
	
	\subsection{Algorithm Convergence Analysis}
	To prove the convergence of the DMM algorithm, the following three conditions should be satisfied \cite{K. Zhou-2022-J}, 1) sufficient decrease condition; 2) a subgradient lower
	bound for the iterates gap; and 3) Kurdyka–Lojasiewicz (KL)
	property \cite{J. Geiping-2018}.

	For the proposed DMM algorithm, the objective function $ f(\mathbf{x}_{n},\mathbf{w}_{n}) $ is firstly transformed into $ f(\mathbf{z}_{n}) $ and then solved by exploiting the MM algorithm twice. Since $ \mathbf{z}_{n}^{(t+1)}=\arg\min\limits_{\mathbf{z}_{n}}u(\mathbf{z}_{n},\mathbf{z}_{n}^{(t)}) $ denotes the optimal solution at $ t+1 $ iteration, according to the principle of the MM algorithm \cite{J. Song-2015}, it is easy to obtain
	\begin{equation}\label{key}
		\begin{split}
			f(\mathbf{x}_{n}^{(t+1)},\mathbf{w}_{n}^{(t+1)})=&f(\mathbf{z}_{n}^{(t+1)})\\
			&\leq u(\mathbf{z}_{n}^{(t+1)},\mathbf{z}_{n}^{(t)})\\
			&\leq u(\mathbf{z}_{n}^{(t)},\mathbf{z}_{n}^{(t)})\\
			&=f(\mathbf{z}_{n}^{(t)})=f(\mathbf{x}_{n}^{(t)},\mathbf{w}_{n}^{(t)}).
		\end{split}
	\end{equation}
	Therefore, the objective function of the optimization problem proposed in this paper is monotonically decreasing. Combining with the relative error condition, we have
	\begin{equation}\label{key}
		\begin{split}
			&\text{dist}(0,\partial f(\mathbf{x}_{n}^{(t+1)},\mathbf{w}_{n}^{(t+1)}))\\
			&\leq c(\parallel\mathbf{x}_{n}^{(t+1)}-\mathbf{x}_{n}^{(t)}\parallel_{2}+\parallel\mathbf{w}_{n}^{(t+1)}-\mathbf{w}_{n}^{(t)}\parallel_{2}),
		\end{split}
	\end{equation}
	where $ c $ is a positive number, and the detailed derivation of (56) can be found in \cite{K. Zhou-2022-J}. Thus, each $ (\mathbf{x}_{n}^{(t)},\mathbf{w}_{n}^{(t)}) $ in the design process is a stationary point of the objective function. \par
	\begin{figure*}[htbp]
		\centering
		\subfigure[]
		{\begin{minipage}{5.9cm}
				\centering
				\includegraphics[width=1\linewidth]{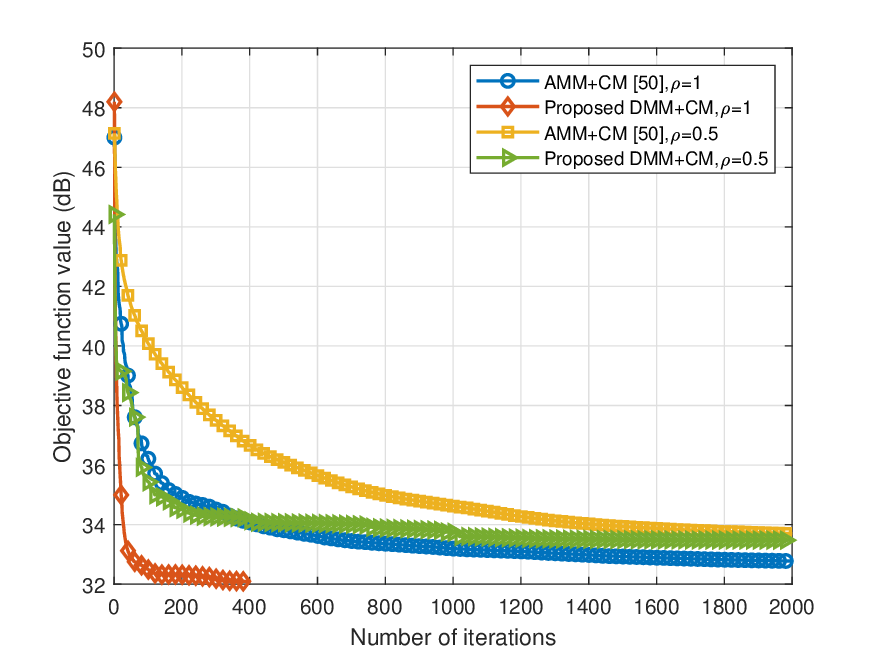}
		\end{minipage}}
		\subfigure[]
		{\begin{minipage}{5.9cm}
				\centering
				\includegraphics[width=1\linewidth]{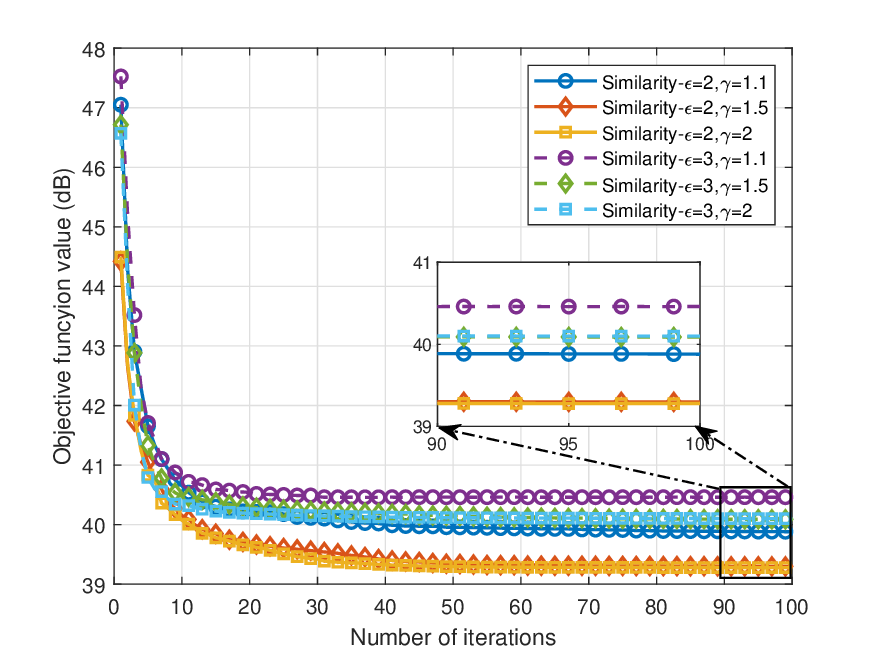}
		\end{minipage}}
		\subfigure[]
		{\begin{minipage}{5.9cm}
				\centering
				\includegraphics[width=1\linewidth]{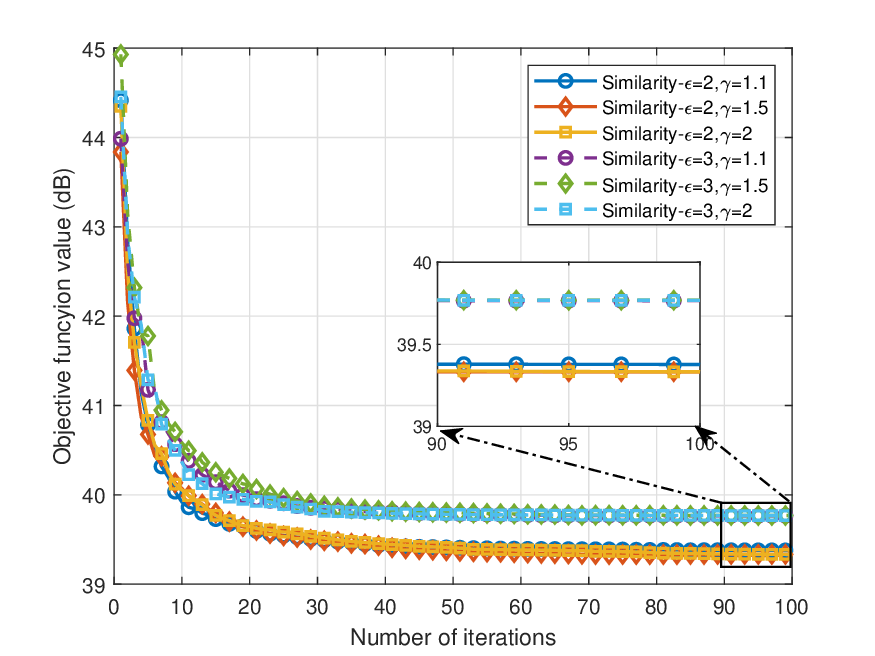}
		\end{minipage}}
		\caption{Convergence performance of the objective function value with respect to the number of iterations. (a) Without communication; (b) 16QAM; (c) QPSK.}
		\label{fig:objective function}
	\end{figure*}
	Further, the objective function $ f(\mathbf{z}_{n}) $ can be converted into a real function with respect to  the real and imaginary parts of $ \mathbf{z}_{n} $. Since all real functions satisfy the KL property \cite{H. B. Chang-2019}, $ \{f(\mathbf{z}_{n}^{(t)})\}$ generated by the DMM algorithm is a Cauchy sequence, which proves the convergence of the DMM algorithm. Specifically, the DMM algorithm does not need to acquine the optimal solution in an alternate iteration way, which accelerates the speed of convergence.
	
	\subsection{Computational Complexity Analysis}
	The computational complexity of the proposed DMM algorithm is mainly related to the number of iterations and the updates of the variable $ \mathbf{z}_{n} $. The computational complexity for updating variable $ \mathbf{z}_{n} $ mainly comes from calculating the $ \mathbf{P}(\mathbf{z}_{n}) $ and the BM to find $ \delta $, 
	where the computation of $\mathbf{P}(\mathbf{z}_{n}) $ involves matrix-vector multiplication. In the calculation of $\mathbf{P}(\mathbf{z}_{n}) $, we introduce the $ \text{FFT} $ operation for rapidly computing $ \mathbf{Q}\mathbf{z}_{n} $ with a computational complexity of $ O(2L\log_{2}(2L)) $. For other matrix-vector multiplication, the computational complexity is $ O(L^{2}) $. The computational complexity of the BM to find the $ \delta $ is $ O(I(L)) $, where $ I $ is the number of iterations of the BM. Therefore, the computation of DMM in one iteration is $ O(2L\log_{2}(2L))+O(L^{2})+O(I(L)) $.

	\section{Numerical evaluations}
	In this section, we demonstrate the performance of the proposed algorithm through numerical and simulation results. Three sets of simulation experiments are designed: 1) the performance analysis of the proposed DMM algorithm; 2) the jamming suppression performance analysis of the proposed ISAC waveform; and 3) the communication symbol error rate (SER) performance analysis of the proposed ISAC waveform. Moreover, the performance curves are obtained through $ 10^{3} $ times Monte-Carlo simulations. Finally, all the numerical simulations are performed on a standard PC with CPU Intel Core i5-12400 and 16 GB RAM.\par
	
	Since the jamming signal is usually accompanied by energy suppression, the jamming-to-signal ratio (JSR) of the jamming signal with respect to the target is defined as \cite{J. Wu-2015} 
	\begin{equation}\label{key}
		\text{JSR}=10\log_{10}\dfrac{|\alpha_{n,J}|^{2}}{|\alpha_{n,T}|^{2}}\;(\text{dB}).
	\end{equation}\par
	In addition, the LPG is defined as the ratio between the radar
	SNR gathered with the mismatched filter
	and the maximum radar SNR obtained by the matched filter \cite{LPG}
	\begin{equation}\label{key}
		\text{LPG}=10\log_{10}\dfrac{\parallel\mathbf{w}_{n}^{H}\mathbf{x}_{n}\parallel_{2}^{2}}{\parallel \mathbf{w}_{n}\parallel^{2}_{2}\parallel \mathbf{x}_{n}\parallel^{2}_{2}}\;(\text{dB}).
	\end{equation}\par
	We assume that the path from the BS to the communication user consists of one line-of-sight (LoS) path and two non-line-of-sight (NLoS) paths, and the delay of the NLoS paths relative to the LoS path is $ 0.5 \mu s $ and $ 0.8 \mu s $, respectively.
	
	\subsection{DMM Algorithm Performance Analysis}
	This subsection first analyzes the convergence of the proposed DMM algorithm. Assume that the ISAC waveform pulse width is $ T_{p}=25.6 \;\mu s$, bandwidth $ B=10\; \text{MHz}$, the sampling time interval is $ t_{s}=0.1\;\mu s $, and the discrete ISAC waveform length is $ L = 256 $. The jamming type is assumed to be the PPRJ, with the sampling time $ T_{L}=4 \;\mu s$ and the number of repeat $ M_{1}=4 $. Moreover, we introduce a weight factor $ \rho=0.4 $  to compromise the optimal sensing and anti-jamming performance. To analyze the LPG caused by the unmatched filter, we set $ a_{\max}=L$ and $ a_{\min}=a_{\max}\cdot 10^{-4} $. Further, we give the weight factors $ \beta_{1}=0.12$ and $\beta_{2}=0.88 $ to compromise the pulse compression peak of the target response and the performance of jamming suppression.\par

	Fig. \ref{fig:objective function} illustrates the curves of the objective function value with respect to the number of iterations for the following three cases: 1) without communication; 2) 16 quadrature amplitude modulation (16QAM); and 3) quadrature phase shift keying (QPSK). It is seen that the proposed DMM algorithm converges quickly whether the communication is considered or not.
	In Fig. \ref{fig:objective function}(a), we compare the proposed DMM algorithm with the AMM algorithm proposed in  \cite{K. Zhou-2022-J}, and we set $ \gamma=1 $ to consider the CM constraint. It is shown that the DMM algorithm converges much faster than the AMM algorithm, since the former does not require alternate iteration. Besides, when we set $ \rho=1 $, i.e., only the sensing function is considered, the convergence speed is quite fast. It is worth highlighting that when we set $ \rho=0.5 $, both the sensing and jamming suppression functions can be achieved, but the convergence speed is decreased. 
	When the communication constraint is considered, as shown in Figs. \ref{fig:objective function}(b) and \ref{fig:objective function}(c), it can be seen that the proposed DMM algorithm converges rapidly to a stationary point under different communication modulations. 
	The penalty parameter $ \epsilon=2 $ corresponds to a lower value of the objective function than in the case of $ \epsilon=3 $, which is because the feasible set of solutions is smaller for the case of $ \epsilon=3 $ than the case of $ \epsilon=2 $.
	Moreover, the results in Fig. \ref{fig:objective function}(b) show that the value of the objective function decreases with the increase of the maximum modulus $ \gamma $ of the ISAC waveform. In particular, the value of the objective function at $ \gamma=1.5 $ is almost the same as the case of $ \gamma=2 $.
	The reason is that the PAPR constraint is weaker than the communication constraint when $ \gamma $ is greater than the maximum modulus $ 3\sqrt{5}/5 $ of 16QAM symbols.
	Fig. \ref{fig:objective function}(c) demonstrates that the objective function value remains almost constant with the increase of $ \gamma $. This is because the QPSK symbols are all constant-modulus, and also the communication constraint is stronger than the PAPR constraint when $ \gamma>1$. Therefore, to obtain the optimal solution under different communication modulations, we set $ \gamma=1.5 $ in the next numerical experiments. Without loss of generality, we assume that QPSK is used for communication modulation in all simulations unless otherwise stated and set the  penalty parameter $ \epsilon=2 $.  \par

	TABLE \ref{time} shows the computational complexity and running time of Fig \ref{fig:objective function}(a), where the maximum number of iterations is set to $ 2\times 10^{3} $.  
	In Fig. \ref{fig:objective function}(a), since we consider the CM constraint, which does not involve the step of the BM, the computational complexity of the proposed DMM algorithm is $O(2L\log_{2}(2L))+O(L^{2}) $.
	As can be seen, since the proposed DMM algorithm introduces FFT to compute the closed-form solution of each sub-problem, it greatly reduces the computational complexity compared to the AMM algorithm. Besides, the experimental results
	shown in TABLE \ref{time} also demonstrate that the proposed DMM algorithm requires less running
	time than that of the AMM algorithm under the same conditions.\par
	
	TABLE \ref{DMMtime} shows the running time of Figs. \ref{fig:objective function}(b) and \ref{fig:objective function}(c) for the proposed DMM algorithm and the BM, where the maximum number of iterations is set to $ 1\times 10^{2} $. We consider different communication modulations and PAPR constraints. Due to the PAPR constraint, the iterative step of the DMM algorithm involves the BM. We can see that the proposed DMM algorithm still has a low runing time and the BM does not cause excessive computation time. This also shows the superiority of the proposed algorithm.
	\begin{table}[H]
		\caption{Comparison of algorithm complexity and running time}\label{time}                    
		\centering                                       
		\begin{tabular}{cccc}                         %c代表该表一共多少列
			\toprule[1.2pt]                                    %第一道横线              
			\multirow{2}{*}{\textbf{Algorithms}}&\multirow{2}{*}{\textbf{Complexity}}& \multicolumn{2}{c}{\textbf{Running time} $ (\text{s}) $} \\
			\Xcline{3-4}{0.5pt}
			& &$ \rho=1 $ &$ \rho=0.5 $\\
			\midrule
			AMM [50]& $ O(L^{3})+O(L^{2}) $&$ 36.53  $& $181.69 $ \\                                     %第二道横线 
			Proposed DMM&$O(2L\log_{2}(2L))+O(L^{2})  $ &$ 0.34  $&$ 38.71  $\\
			\bottomrule[1.2pt]                                   %第三道横线
		\end{tabular}
	\end{table}\par
	\begin{table}[H]
		\caption{Running time of the proposed DMM algorithm and the BM}\label{DMMtime}                    
		\centering                                       
		\begin{tabular}{ccccc}                         %c代表该表一共多少列
			\toprule[1.2pt]                                    %第一道横线              
			\multirow{2}{*}{\textbf{Modulations}}&\multirow{2}{*}{\textbf{Parameter}}&
			\multicolumn{3}{c}{\textbf{Running time: DMM/BM} $ (\text{s}) $}\\ 
			\Xcline{3-5}{0.5pt}
			& &$ \gamma=1.1 $ &$  \gamma=1.3 $&$  \gamma=1.5 $\\
			\midrule
			\multirow{2}{*}{16QAM}& $\epsilon=2 $&$ 2.78/0.15  $& $2.69/0.15 $ &$2.73/0.15 $\\         %第二道横线 
			& $ \epsilon=3$&$ 2.73/0.15  $& $2.72/0.15 $ &$2.95/0.15 $\\
			\multirow{2}{*}{QPSK}&$ \epsilon=2  $ &$ 2.71/0.15  $&$ 2.73/0.16  $&$2.72/0.15 $\\
			& $\epsilon=3 $&$ 2.79/0.16 $& $2.78/0.15 $ &$2.82/0.15 $\\
			\bottomrule[1.2pt]                                   %第三道横线
		\end{tabular}
	\end{table}\par
	
	\begin{figure}
		\centering
		\includegraphics[width=0.9\linewidth]{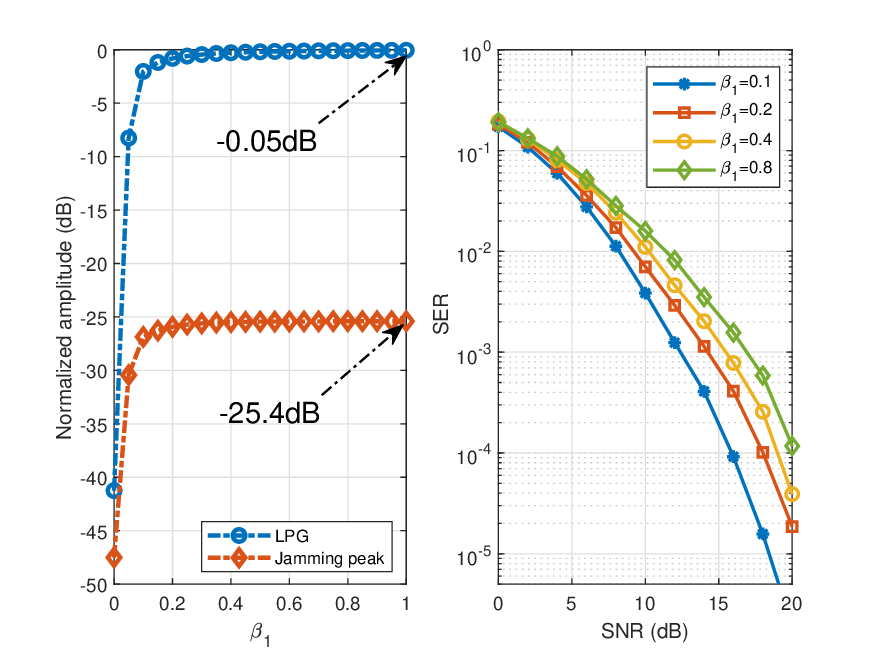}
		\caption{Evaluation of the LPG, jamming peak and communication SER at different $ \beta_{1} $.}
		\label{fig:betalpgjammingpeakser}
	\end{figure}
	
	\begin{figure}[h]
		\centering
		\includegraphics[width=0.9\linewidth]{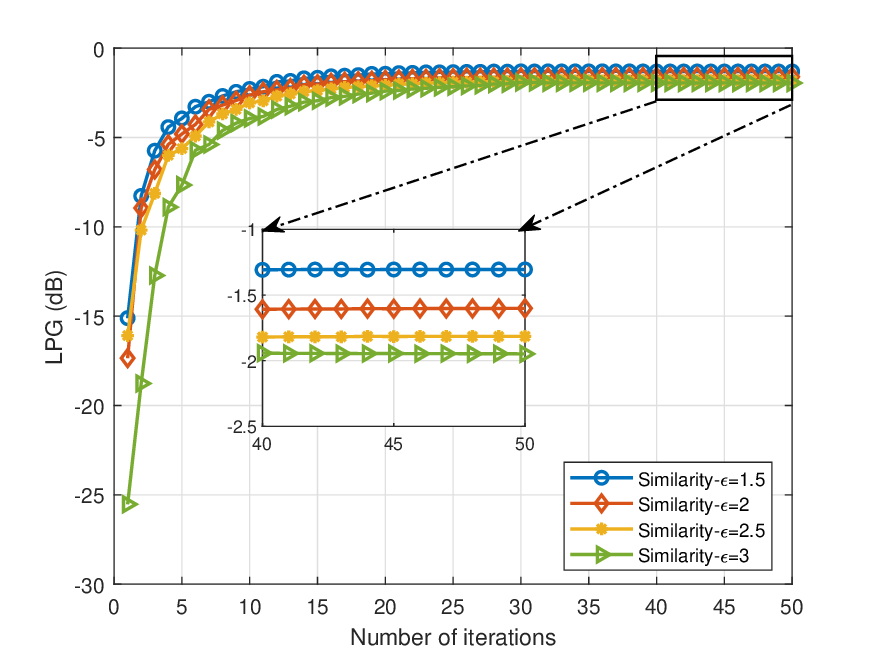}
		\caption{ Evaluation of the LPG curves with the number of iterations.}
		\label{fig:LPG}
	\end{figure}

	To analyze the impact of parameter $\beta_{1}$, Fig. \ref{fig:betalpgjammingpeakser} illustrates the variation trend of LPG, jamming peak and communication SER with different values of $\beta_{1} $. It can be seen that the LPG increases with the growth of $\beta_{1} $. On the contrary, the jamming peak is suppressed when the $\beta_{1} $ decreases. However, LPG and communication SER are coupled to each other, and in Fig. \ref{fig:betalpgjammingpeakser} one can see that an increase in $ \beta_{1} $ leads to an increase in SER.
	Thus, the weight $ \beta_{1} $ should be reasonably chosen to precisely control LPG, jamming peak and communication SER. By weighing between the target peak, the jamming peak and communication SER, we set $ \beta_{1}=0.12 $ in subsequent simulations.

	Furthermore, in Fig. \ref{fig:LPG}, we illustrate the value of LPG with respect to the number of iterations under different values of the penalty parameter $\epsilon $. It is shown that the LPG increases with the increase of the number of iterations, but it will gradually stabilize. Besides, the LPG can be further decreased by reducing the values of penalty parameter $ \epsilon $. 
	\begin{figure*}[htbp]
		\centering
		\subfigure[]
		{\begin{minipage}{5.9cm}
				\centering
				\includegraphics[width=1\linewidth]{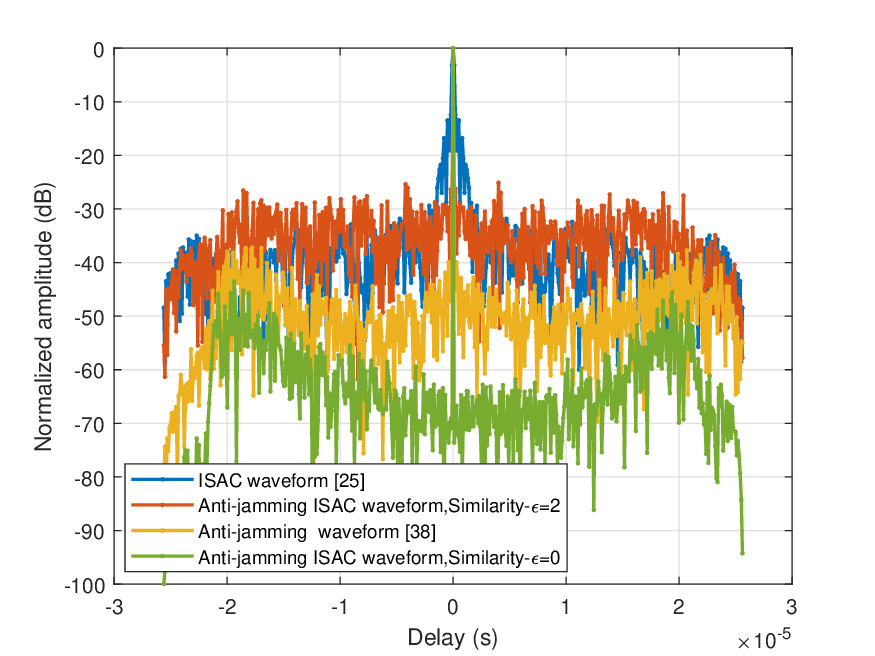}
		\end{minipage}}
		\subfigure[]
		{\begin{minipage}{5.9cm}
				\centering
				\includegraphics[width=1\linewidth]{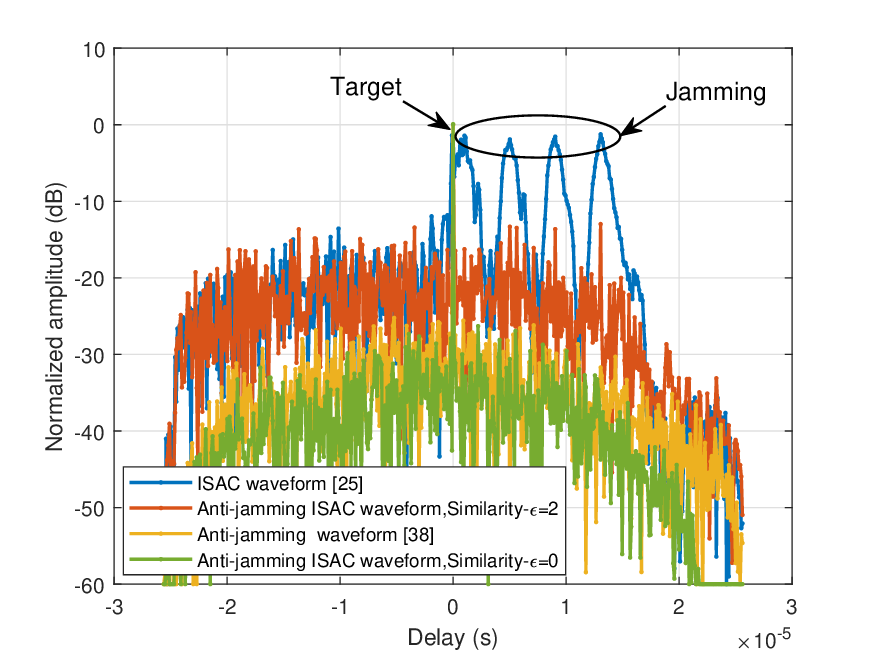}
		\end{minipage}}
		\subfigure[]
		{\begin{minipage}{5.9cm}
				\centering
				\includegraphics[width=1\linewidth]{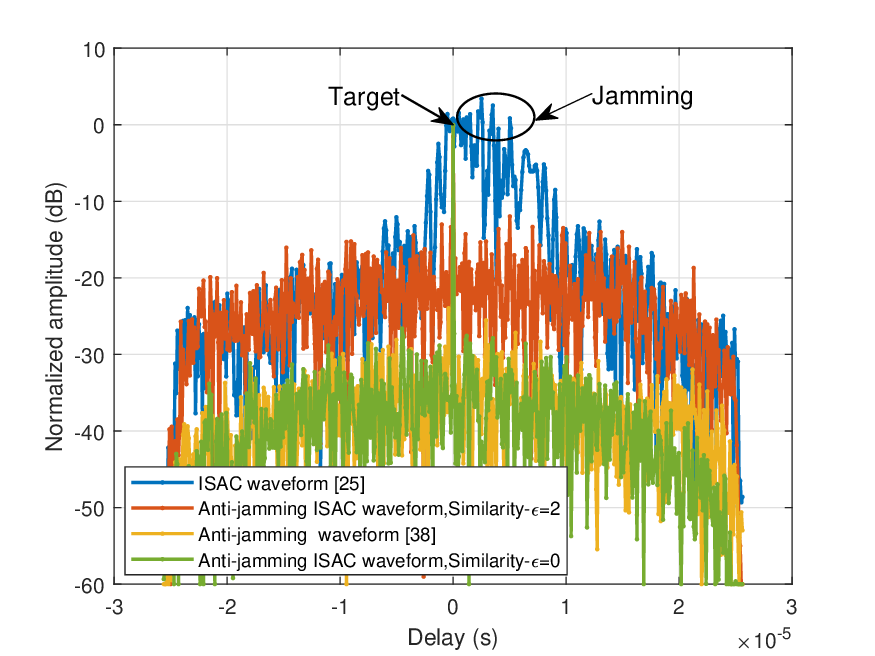}
		\end{minipage}}
		\caption{Sensing and jamming suppression performance comparison ($ \gamma=1.5 $). (a) Pulse compression results without jamming; (b) Jamming suppression performance for PPRJ scenario (JSR = $ 15 $ dB); (c) Jamming suppression performance for RRJ scenario (JSR = $ 15 $ dB).}
		\label{fig:waveform}
	\end{figure*}

	\subsection{Performance of Sensing and Jamming Suppression}

	In this section, simulation results are performed to verify the anti-jamming effect and sensing performance of the proposed ISAC waveform. We assume that the weight factor is $ \rho=0.4 $,
	and the algorithm convergence threshold is $ \eta=1\times 10^{-5} $. Simulation and jamming-related parameters are given in TABLE \ref{radar}. Moreover, two benchmarks, an anti-jamming waveform proposed in  \cite{M. Ge-2021} using the decoupled alternating direction penalty method (DCADPM) algorithm and an IASC waveform proposed in \cite{F. Liu-2018-2}, are adopted for comprison. Particularly, in the case of a single-antenna, the ISAC waveform based on trade-off design proposed in \cite{F. Liu-2018-2} is formulated as
	\begin{equation}\label{key}
		\begin{aligned}
			\min_{\mathbf{x}_{n}}\;&\rho\parallel \mathbf{H}\mathbf{x}_{n}-\mathbf{s}_{n}\parallel_{2}^{2}+(1-\rho)\parallel \mathbf{x}_{n}-\mathbf{x}_{rad,n}\parallel_{2}^{2}&\\
			\mbox{s.t.}\;\;&\parallel\mathbf{x}_{n}\parallel_{2}^{2}=L,&\\
			& \text{PAPR}(\mathbf{x}_{n})\leq\gamma^{2},
		\end{aligned}
	\end{equation}
	where  $ \mathbf{x}_{rad,n} $ is the given optimal radar waveform, which is assumed as the linear frequency modulation (LFM) in the following. The pulse width and bandwidth of the LFM signal are consistent with the ISAC waveform.\par
	
	To simplify the notation, in the following, ``Anti-jamming ISAC waveform" denotes the proposed design in this work, ``ISAC waveform" represents the waveform obtained by \cite{F. Liu-2018-2} and ``Anti-jamming waveform" is acquired by \cite{M. Ge-2021}. 
	\begin{table}[h]
		\caption{Simulation parameters}\label{radar}                    %标题
		\centering                                       %把表居中
		\begin{tabular}{cccc}                         %7个c代表该表一共7列  ，内容全部居中
			\toprule[1.2pt]                                    %第一道横线              
			\multirow{3}{*}{\textbf{Parameters}} &\multicolumn{3}{c}{\textbf{Value}} \\
			\Xcline{2-4}{0.5pt}
			&\multirow{2}{*}{\textbf{Sensing}} &\multicolumn{2}{c}{\textbf{Jamming}} \\
			\Xcline{3-4}{0.5pt}
			& &PPRJ&RRJ\\
			\midrule 
			Pulse width $ T_{p}$& $ 25.6 \;\mu s$&$ - $ &$ - $ \\                                     %第二道横线 
			Bandwidth $ B $ & $ 10$ MHz&$ - $ &$ - $\\
			Pulse repetition frequency $ f_{r} $ &$5 $ kHz&$ - $ &$ - $\\
			Sampling time interval $ t_{s} $ & $ 0.1\;\mu s $ &$ 0.1\;\mu s $&$ 0.1\;\mu s $\\
			Jamming sampling time $ T_{L} $ &$ - $ &$ 4\;\mu s $ &$ 1\;\mu s $\\
			Jamming sampling interval $ T_{s} $&$ - $ &$ - $ &$6. 4\;\mu s $\\
			The number of repeats &$ - $ & $ 4 $& $ 5 $\\
			Jammer relative target echo delay&$ - $ &$ 1 \;\mu s$ &$ 1 \;\mu s$\\
			Jamming-to-signal ratio (JSR)&$ - $ &$ 15 $ dB & $ 15 $ dB\\
			Signal-to-noise ratio (SNR)&$ 10 $ dB&$ - $ &$ - $\\
			\bottomrule[1.2pt]                                   %第三道横线
		\end{tabular}
	\end{table}\par

	Fig. \ref{fig:waveform}(a) presents the pulse compression results without considering jamming. To ensure superior sensing performance, we set $ \rho=0.1 $ for the ISAC waveform, and set $ \rho=0.4 $ for the Anti-jamming waveform. As can be seen, the peak sidelobe level ratio (PSLR) \cite{LPG} of the ISAC waveform and the Anti-jamming waveform are $ -13.36 $ dB, $ -37.29 $ dB, respectively. When $ \epsilon=2 $, i.e., communication function is considered, the PSLR of the proposed Anti-jamming ISAC waveform  is $ -23.51 $ dB. Compared to the ISAC waveform, the proposed Anti-jamming ISAC waveform has shown significant improvements in terms of PSLR performance. However, it is important to note that the ISAC waveform uses the LFM as the optimal radar waveform, which presents good ambiguity properties while performing well in distinguishing moving targets \cite{M. A. Richards-2014}.
	Additionally, when $ \epsilon=0 $, i.e., without communication, the PSLR of the proposed Anti-jamming ISAC waveform is $ -43.52 $ dB. In conclusion, though the Anti-jamming ISAC waveform dispalys sensing performance degradation compared with the Anti-jamming waveform in the ISAC scenario, it shows superior PSLR property without considering the communication function.
	Further, the pulse compression results in the presence of PPRJ and RRJ are shown in Figs. \ref{fig:waveform}(b) and \ref{fig:waveform}(c), respectively.
	It is seen that when transmits the ISAC waveform, the false targets and the true target appear simultaneously, resulting an increase of the radar false alarm probability. On the contrary, when using the proposed  Anti-jamming ISAC waveform and the Anti-jamming waveform, both the PPRJ and RRJ can be effectively filtered and suppressed, realizing the improvement of sensing performance. For the sake of analysis, the jamming type is assumed to be the PPRJ in the subsequent simulations.\par
	
	\begin{figure}[htbp]
		\centering
		\subfigure[]
		{\begin{minipage}{4.3cm}
				\centering
				\includegraphics[width=1.1\linewidth]{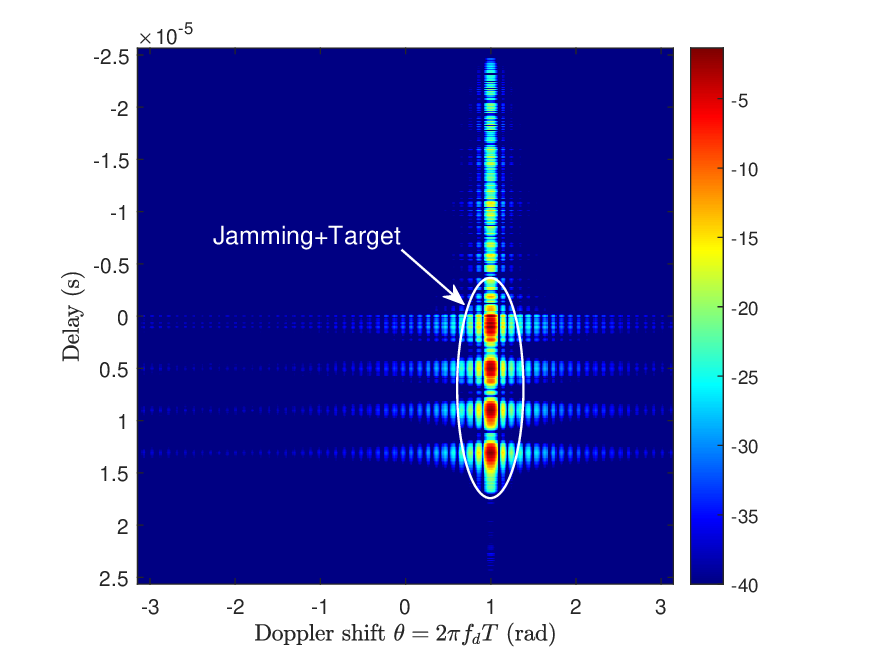}
		\end{minipage}}
		\subfigure[]
		{\begin{minipage}{4.3cm}
				\centering
				\includegraphics[width=1.1\linewidth]{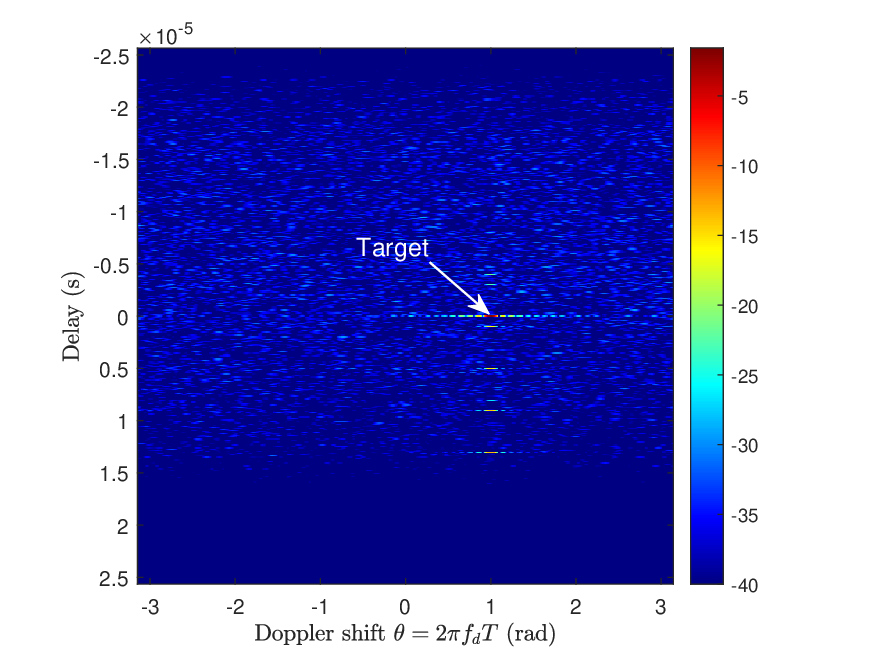}
		\end{minipage}}
		\subfigure[]
		{\begin{minipage}{4.3cm}
				\centering
				\includegraphics[width=1.1\linewidth]{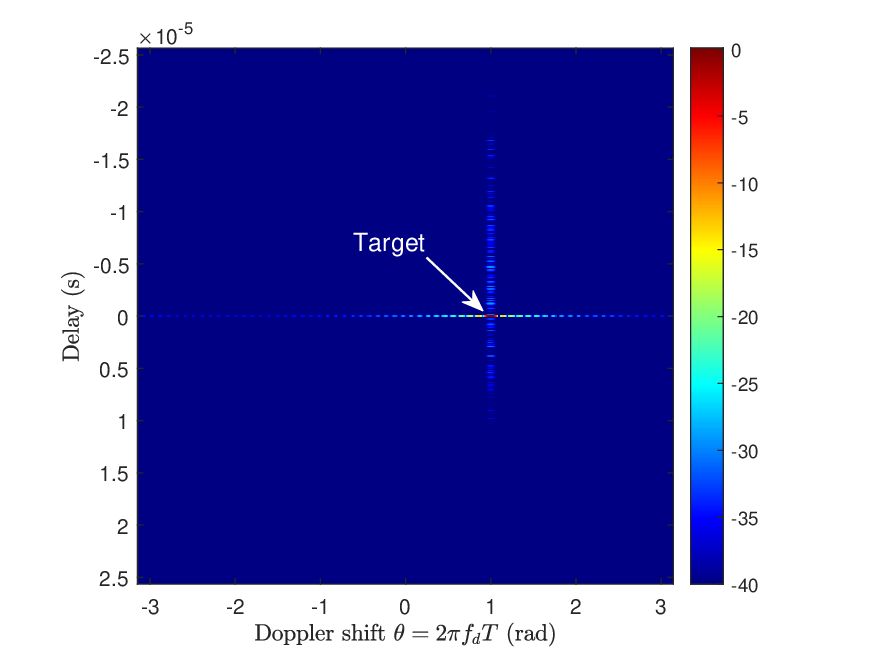}
		\end{minipage}}
		{\begin{minipage}{4.3cm}
				\centering
				\includegraphics[width=1.1\linewidth]{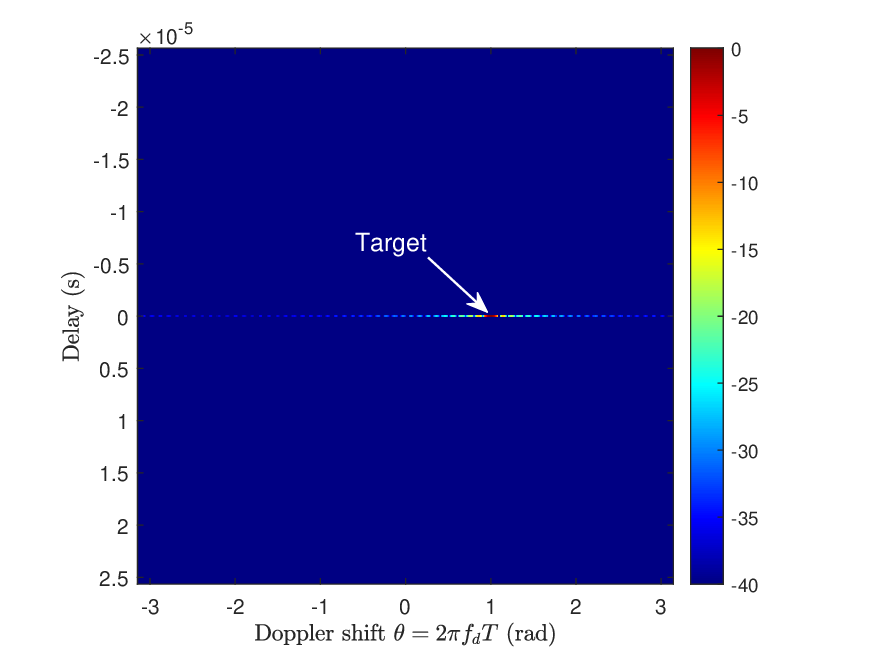}
		\end{minipage}}
		\caption{The Delay-Doppler images for PPRJ scenario ($ N = 64 $, JSR = $ 15 $ dB, $ \gamma=1.5  $). (a) ISAC waveform \cite{F. Liu-2018-2}; (b) Anti-jamming ISAC waveform ($ \epsilon=2 $); (c) Anti-jamming waveform \cite{M. Ge-2021}; (d) Anti-jamming ISAC waveform  ($ \epsilon=0 $).}
		\label{fig:RD}
	\end{figure}
	
	Fig. \ref{fig:RD} demonstrates the Delay-Doppler images in the PPRJ scenario. We assume that the normalized Doppler shift $ \theta=1 \;\text{rad}$ of the target and the jamming. It can be seen that when the ISAC waveform is used, the real target is obscured by jamming. Nevertheless, false targets generated by PPRJ are effectively suppressed by using the proposed Anti-jamming ISAC waveform and the Anti-jamming waveform. It is worth noting that the proposed method has a higher target SNR gain on the radar the Delay-Doppler image compared to the DCADPM algorithm proposed in \cite{M. Ge-2021} when communication is not considered, which further proves the effectiveness of the proposed method.
	
	\begin{Remark}
		
		In Fig. \ref{fig:RD}, the filter group $ \{\mathbf{w}_{n}e^{jn\theta_{1}},\mathbf{w}_{n}e^{jn\theta_{2}},\cdots,$ $ \mathbf{w}_{n}e^{jn\theta_{M}}\}$, $\theta_{i}=-\pi+2\pi (i-1)/(M-1),i=1,2,\cdots,M$ is applied to the radar receive signal to detect the target, where $ M=201 $. The Doppler resolution of the Delay-Doppler maps is higher for larger  $ M  $. In addition, the target delay is obtained by coherent accumulation sum of each pulse within a coherent processing interval. Therefore, increasing the number of pulses $ N $ could further reduce the sidelobes of the Delay-Doppler images and improve the resolution.
	\end{Remark}
	\begin{figure}[h]
		\centering
		\includegraphics[width=0.9\linewidth]{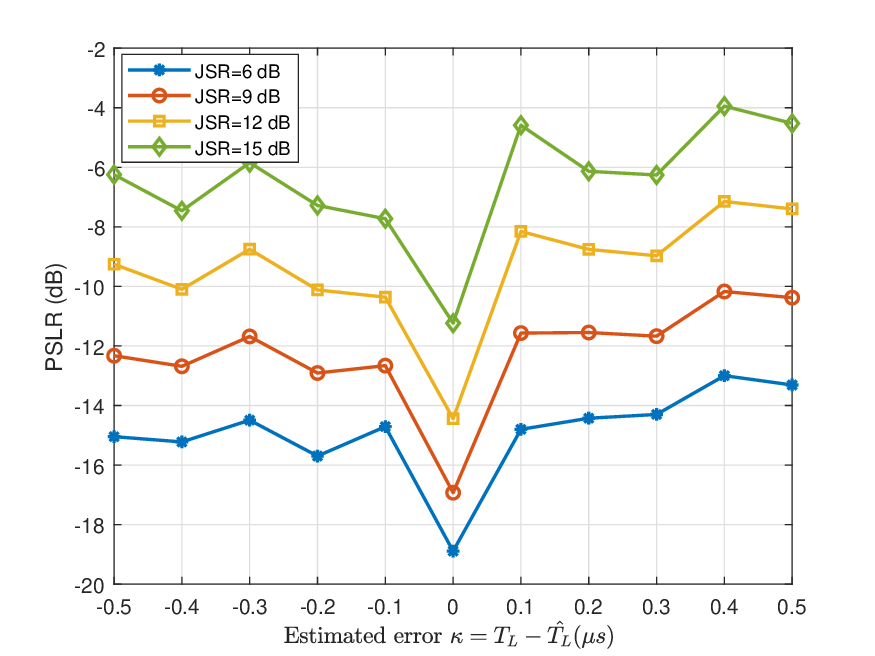}
		\caption{The variation trends of the PSLR under different error $ \kappa $ of the jamming sampling time ($ \epsilon=2 $).}
		\label{fig:PSLR}
	\end{figure}

	Moreover, the cognitive-based method usually does not exactly obtain the relevant parameters of the jamming in practice, resulting in appearing error in estimated jamming parameters. To analyze the effect of jamming cognitive errors on the performance of the proposed method, 	
	Fig. \ref{fig:PSLR} depicts the trend of the PSLR with respect to the estimated error $ \kappa $ of the jamming sampling time, where the estimated jamming sampling time $ \hat{T}_{L}=4\mu s $ and $ \kappa=T_{L}-\hat{T}_{L} $, $ T_{L} $ is actual jamming sampling time. In Fig. \ref{fig:PSLR}, it can be seen that the PSLR is lowest when the estimated error $ \kappa =0 \mu s $, i.e., the actual $ {T}_{L} $ is the same as the estimated $ \hat{T}_{L} $. Not surprisingly, the PSLR performance is deteriorated in the case of appearing estimation error. However, although inaccurate jamming information leads to a degradation of the jamming suppression performance, it also ensures the target detection in the case of a low JSR.
	\par
	\begin{figure}[h]
		\centering
		\includegraphics[width=0.9\linewidth]{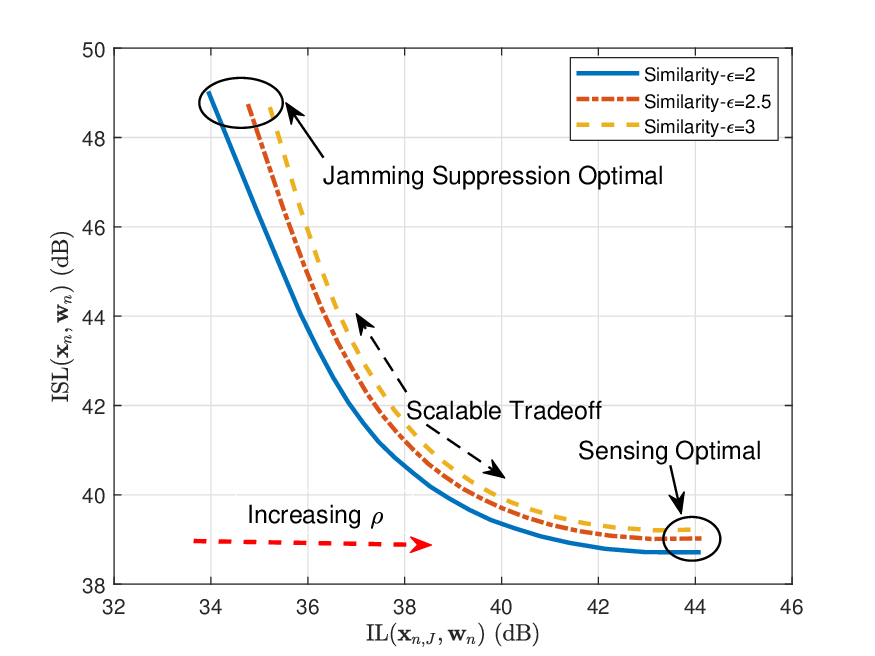}
		\caption{Trade-off between $ \text{ISL}(\mathbf{x}_{n},\mathbf{w}_{n}) $ and  $ \text{IL}(\mathbf{x}_{n,J},\mathbf{w}_{n}) $.}
		\label{fig:ISL}
	\end{figure}
	In Fig. \ref{fig:ISL}, we aim at explicitly showing the trade-offs
	between the sensing and anti-jamming performance with different values of the penalty parameter $ \epsilon $. We use the $ \text{ISL}(\mathbf{x}_{n},\mathbf{w}_{n}) $ and the $ \text{IL}(\mathbf{x}_{n,J},\mathbf{w}_{n}) $ as the metrics of sensing and jamming suppression performance, respectively.
	As can be seen, the sensing and anti-jamming performance increase
	with the decrease of $ \epsilon $, indicating that the $ \text{ISL}(\mathbf{x}_{n},\mathbf{w}_{n}) $ and $ \text{IL}(\mathbf{x}_{n,J},\mathbf{w}_{n}) $ can be further minimized by sacrificing the communication performance.
	Besides, there is a trade-off between the jamming suppression and sensing. 
	It is evident from the optimization problem $ \mathcal{P}_{0} $ that a larger $ \rho $ imposes a greater weight on the ISL,  whereby the sensing performance is more enhanced than the jamming suppression performance. Therefore, a trade-off between the jamming suppression and sensing can be achieved by reasonably adjusting  $\rho$ in practice. \par

	\subsection{Communication Performance}
	To evaluate the communication performance of the ISAC system, the communication SER is evaluated in this section. 
	
	\begin{figure}[htbp]
		\centering
		\subfigure[]
		{\begin{minipage}{8cm}
				\centering
				\includegraphics[width=1\linewidth]{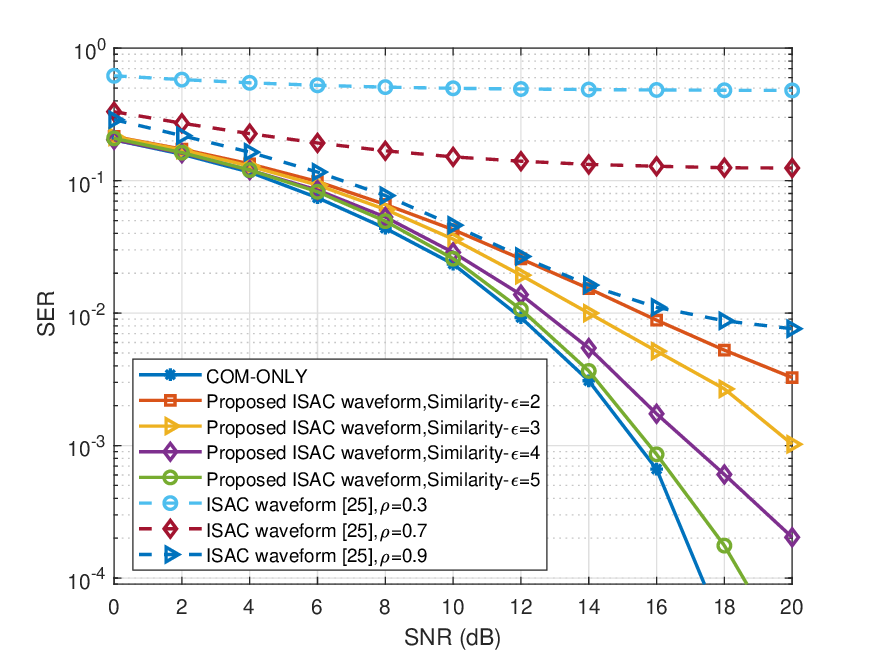}
		\end{minipage}}
		\subfigure[]
		{\begin{minipage}{8cm}
				\centering
				\includegraphics[width=1\linewidth]{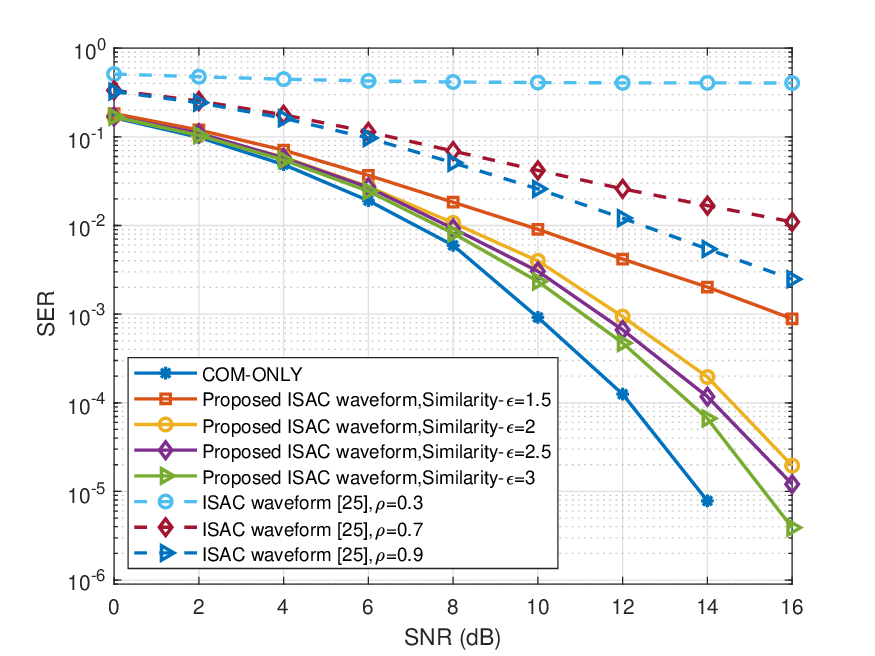}
		\end{minipage}}
		\caption{The SER comparison for different $ \epsilon $. (a) SER performance with 16QAM modulation; (b) SER performance with QPSK modulation.}
		\label{fig:SER}
	\end{figure}
	
	Fig. \ref{fig:SER}(a) and Fig. \ref{fig:SER}(b) respectively display the SER of the 16QAM and QPSK modulations. For comparison, we consider the case of the communication only (COM-ONLY), and the ISAC waveform proposed in \cite{F. Liu-2018-2}.
	It can be seen that as the penalty parameter $ \epsilon $ increases, the performance of proposed method gradually approaches the COM-ONLY.
	The reason is that according to the optimization problem (15), a larger $ \epsilon $ indicates more resources are allocated to optimize the communication performance, resulting in a better SER performance, but one loses the performance of sensing and jamming suppression as shown in Fig. \ref{fig:ISL}. Thus, an appropriate $ \epsilon $ should be selected according to the actual requirements. 
	Moreover, we see that the SER of the ISAC waveform proposed in \cite{F. Liu-2018-2} can be further reduced by increasing the weighting factor $ \rho $, but the communication SER is still higher than that of the proposed method.

	\par
	\begin{figure}[h]
		\centering
		\includegraphics[width=0.9\linewidth]{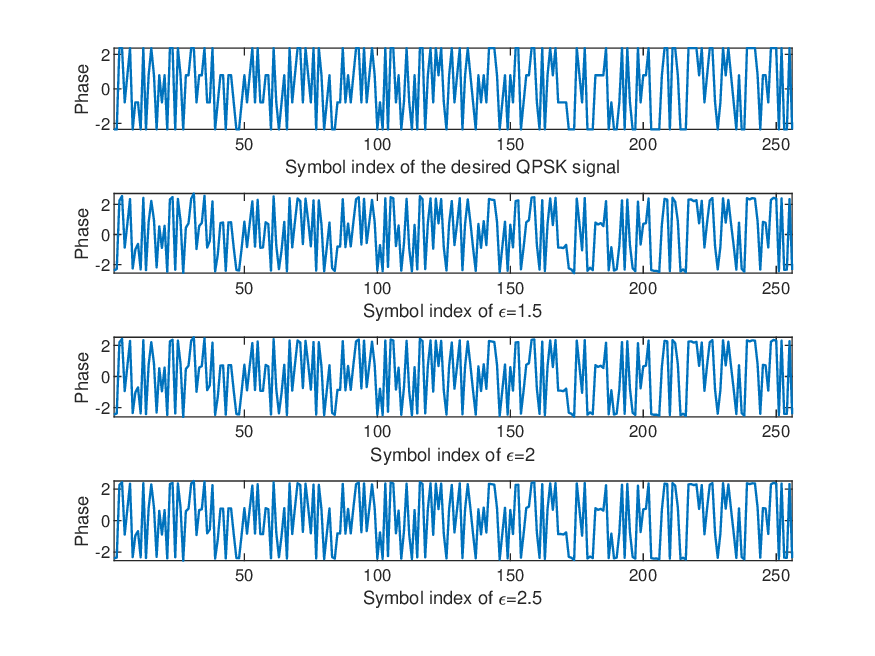}
		\caption{The phases of the desired QPSK signal and the designed ISAC waveform at different $ \epsilon $.}
		\label{fig:Phase}
	\end{figure}

	As the final part of this section, the impact of the similarity constraint on the transmitted waveform is investigated. We assume that the ISAC system sends a QPSK signal to the communication receiver, and the amplitude of the desired QPSK signal is 1 and the symbol bits are randomly generated. Fig. \ref{fig:Phase} shows the phases of the desired QPSK signal and the designed ISAC waveform at different $ \epsilon $.  We can see that the phases of the designed ISAC waveform approximate the desired communication signal at different $ \epsilon $, and high penalty parameter $ \epsilon $ result in a better approximation of the phases of the desired communication signal.

	\section{Conclusions}
	In this paper, the joint design of the transmitted waveform and the receive filter for the ISAC system has been proposed in the presence of the SDMJ. 
	Specifically, we have developed
	an objective optimization criterion through merging the ISL of the transmitted waveform and the IL of the jamming,
	which enables a flexible performance trade-off between the sensing and the jamming suppression. The communication performance of the ISAC system is further ensured by considering the similarity constraint.
	To solve the formulated non-convex problem, we have first developed an efficient DMM algorithm based on the MM framework, and then an acceleration algorithm based on the square iteration method has been used to speed up  the convergence rate. Moreover, we have derived a closed-form solution to the proposed problem, and proved the convergence of the proposed solution. The results show that the proposed DMM algorithm has a faster convergence rate and lower complexity compared to the traditional MM algorithm based on alternating iterations.
	It is shown that the proposed scheme can effectively suppress PPRJ and RRJ, and also exhibit superior sensing and communication performance. Note that the proposed method cannot control the value of the LPG accurately. Besides, the pulse compression performance gets deteriorated for large Doppler shift.
	As a future work, Doppler-resilient ISAC waveforms may be designed.

	\appendices
	\section{proof of (20)}
	
	$ Proof $: By calculating, we have
	\begin{equation}\label{key}
		\begin{split}
			&\rho\sum_{k\in\bm{\Omega}}|\mathbf{z}_{n}^{H}\widetilde{\mathbf{U}}_{k}\mathbf{z}_{n}|^{2}+(1-\rho)\sum_{k\in\bm{\Omega}_{J}}|\mathbf{z}_{n,J}^{H}\widetilde{\mathbf{U}}_{k}\mathbf{z}_{n}|^{2}\\
			=&\rho\sum_{k\in\bm{\Omega}}|\mathbf{Tr}(\widetilde{\mathbf{U}}_{k}\mathbf{Z}_{n})|^{2}+(1-\rho)\sum_{k\in\bm{\Omega}_{J}}|\mathbf{Tr}(\widetilde{\mathbf{U}}_{J,k}\mathbf{Z}_{n})|^{2}\\
			=&\rho\sum_{k\in\bm{\Omega}}\text{vec}(\mathbf{Z}_{n})^{H}\text{vec}(\widetilde{\mathbf{U}}_{k})\text{vec}(\widetilde{\mathbf{U}}_{k})^{H}\text{vec}(\mathbf{Z}_{n})\\
			&+(1-\rho)\sum_{k\in\bm{\Omega}_{J}}\text{vec}(\mathbf{Z}_{n})^{H}\text{vec}(\widetilde{\mathbf{U}}_{J,k})\text{vec}(\widetilde{\mathbf{U}}_{J,k})^{H}\text{vec}(\mathbf{Z}_{n})\\
			=&\text{vec}(\mathbf{Z}_{n})^{H}\left[\rho\sum_{k\in\bm{\Omega}}\text{vec}(\widetilde{\mathbf{U}}_{k})\text{vec}(\widetilde{\mathbf{U}}_{k})^{H}\right]\text{vec}(\mathbf{Z}_{n})\\
			&+\text{vec}(\mathbf{Z}_{n})^{H}\left[(1-\rho)\sum_{k\in\bm{\Omega}_{J}}\text{vec}(\widetilde{\mathbf{U}}_{J,k})\text{vec}(\widetilde{\mathbf{U}}_{J,k})^{H}\right]\text{vec}(\mathbf{Z}_{n}),
		\end{split}
	\end{equation}
	where $\mathbf{Z}_{n}=\mathbf{z}_{n}\mathbf{z}_{n}^{H}$, and
	\begin{equation}\label{key}
		\widetilde{\mathbf{U}}_{J,k}=\begin{bmatrix}
			\mathbf{O}&\mathbf{J}^{H}\mathbf{U}_{k}\\
			\mathbf{O}&\mathbf{O}
		\end{bmatrix}.
	\end{equation}
	Then, it is easy to verify that the objective function of the $ \mathcal{P}_{1} $ can be rewritten as (20). The proof is completed. $ \hfill\blacksquare $

	\section{proof of Lemma 2}
	$ Proof $: It is easy to see that the set of vectors $ \{\text{vec}(\widetilde{\mathbf{U}}_{k})|k\in\bm{\Omega}\}$ are mutually orthogonal. For $ k\in\bm{\Omega}$, we have
	\begin{equation}\label{key}
		\begin{split}
			\mathbf{A}\text{vec}(\widetilde{\mathbf{U}}_{k})&=\sum_{i=1-L,i\neq 0}^{L-1}\text{vec}(\widetilde{\mathbf{U}}_{i})\text{vec}(\widetilde{\mathbf{U}}_{i})^{H}\text{vec}(\widetilde{\mathbf{U}}_{k})\\
			&=\text{vec}(\widetilde{\mathbf{U}}_{k})\text{vec}(\widetilde{\mathbf{U}}_{k})^{H}\text{vec}(\widetilde{\mathbf{U}}_{k})\\
			&=(L-|k|)\text{vec}(\widetilde{\mathbf{U}}_{k}).
		\end{split}
	\end{equation}
	Thus, $ (L-|k|), k\in\bm{\Omega}$, are the nonzero eigenvalues with corresponding eigenvectors $ \text{vec}(\widetilde{\mathbf{U}}_{k}),k\in\bm{\Omega} $. Then, the maximum eigenvalue of $ \mathbf{A} $ is given by $\max\limits_{k}\{(L-|k|)k\in\bm{\Omega}\}=L-1 $.
	
	For the same reason, we have
	\begin{equation}\label{key}
		\begin{split}
			\mathbf{B}\text{vec}(\widetilde{\mathbf{U}}_{J,k})&=\sum_{i=1-L}^{L-1}\text{vec}(\widetilde{\mathbf{U}}_{J,i})\text{vec}(\widetilde{\mathbf{U}}_{J,i})^{H}\text{vec}(\widetilde{\mathbf{U}}_{J,k})\\
			&=\text{vec}(\widetilde{\mathbf{U}}_{J,k})\text{vec}(\widetilde{\mathbf{U}}_{J,k})^{H}\text{vec}(\widetilde{\mathbf{U}}_{J,k})\\
			&=[\text{vec}(\widetilde{\mathbf{U}}_{J,k})^{H}\text{vec}(\widetilde{\mathbf{U}}_{J,k})]\text{vec}(\widetilde{\mathbf{U}}_{J,k})\\
			&=[\text{vec}(\mathbf{J}^{H}\mathbf{U}_{k})^{H}\text{vec}(\mathbf{J}^{H}\mathbf{U}_{k})]\text{vec}(\widetilde{\mathbf{U}}_{J,k}).
		\end{split}
	\end{equation}
	Therefore, $ \text{vec}(\mathbf{J}^{H}\mathbf{U}_{k})^{H}\text{vec}(\mathbf{J}^{H}\mathbf{U}_{k}) $ are the nonzero eigenvalues with corresponding eigenvectors $ \text{vec}(\widetilde{\mathbf{U}}_{J,k}),k\in\bm{\Omega}_{J} $. Then, the maximum eigenvalue of $ \mathbf{B} $ is given by $\max\limits_{k}\{\text{vec}(\mathbf{J}^{H}\mathbf{U}_{k})^{H}\text{vec}(\mathbf{J}^{H}\mathbf{U}_{k})|k\in\bm{\Omega}_{J}\} $. The proof is completed. $ \hfill\blacksquare $\par

	\section{proof of (28)}
	$ Proof $: According to (27), we have
	\begin{equation}\label{key}
		\begin{split}
			&\text{Re}\left\{\text{vec}(\mathbf{Z}_{n})^{H}(\mathbf{W}-\lambda_{u}\mathbf{I})\text{vec}(\mathbf{Z}_{n}^{(t)})\right\}\\
			=&\rho\text{Re}\left\{\text{vec}(\mathbf{Z}_{n})^{H} \mathbf{A}\text{vec}(\mathbf{Z}_{n}^{(t)})\right\}\\
			&+(1-\rho)\text{Re}\left\{\text{vec}(\mathbf{Z}_{n})^{H}\mathbf{B}\text{vec}(\mathbf{Z}_{n}^{(t)})\right\}\\
			&-\text{Re}\left\{\text{vec}(\mathbf{Z}_{n})^{H}(\lambda_{u}\mathbf{I})\text{vec}(\mathbf{Z}_{n}^{(t)})\right\}.
		\end{split}
	\end{equation}\par
	Firstly, according to (21), we can obtain
	\begin{equation}\label{key}
		\begin{split}
			&\text{Re}\left\{\text{vec}(\mathbf{Z}_{n})^{H} \mathbf{A}\text{vec}(\mathbf{Z}_{n}^{(t)})\right\}\\
			=&\text{Re}\left\{\text{vec}(\mathbf{Z}_{n})^{H}\left(\sum_{k\in\bm{\Omega}}\text{vec}(\widetilde{\mathbf{U}}_{k})\text{vec}(\widetilde{\mathbf{U}}_{k})^{H}\right)\text{vec}(\mathbf{Z}_{n}^{(t)})\right\}\\
			=&\text{Re}\left\{\sum_{k\in\bm{\Omega}}(\mathbf{Tr}(\widetilde{\mathbf{U}}_{-k}\mathbf{Z}_{n}^{(t)})\mathbf{Tr}(\widetilde{\mathbf{U}}_{k}\mathbf{Z}_{n}))\right\}\\
			=&\text{Re}\left\{\mathbf{Tr}\left(\sum_{k\in\bm{\Omega}}C_{\mathbf{w}_{n},\mathbf{x}_{n}}^{(t)}(-k)\widetilde{\mathbf{U}}_{k}\mathbf{Z}_{n}\right)\right\}\\
			=&\text{Re}\left\{\mathbf{Tr}(\mathbf{Q}_{1}\mathbf{Z}_{n})\right\},
		\end{split}
	\end{equation}
	where 
	\begin{equation}\label{key}
		\mathbf{Q}_{1}
		=\begin{bmatrix}
			\mathbf{O}&\bm{\Phi}\\
			\mathbf{O}&\mathbf{O}
		\end{bmatrix},\;\bm{\Phi}=\sum\limits_{k\in\bm{\Omega}}C_{\mathbf{w}_{n},\mathbf{x}_{n}}^{(t)}(-k)\mathbf{U}_{k}.
	\end{equation}\par
	
	Similarly, by (22) we can compute
	\begin{equation}\label{key}
		\begin{split}
			&\text{Re}\left\{\text{vec}(\mathbf{Z}_{n})^{H}\mathbf{B}\text{vec}(\mathbf{Z}_{n}^{(t)})\right\}\\
			=&\text{Re}\left\{\text{vec}(\mathbf{Z}_{n})^{H}\left(\sum_{k\in\bm{\Omega}_{J}}\text{vec}(\widetilde{\mathbf{U}}_{J,k})\text{vec}(\widetilde{\mathbf{U}}_{J,k})^{H}\right)\text{vec}(\mathbf{Z}_{n}^{(t)})\right\}\\
			=&\text{Re}\left\{\sum_{k\in\bm{\Omega}_{J}}\mathbf{Tr}(\widetilde{\mathbf{U}}_{J,-k}\mathbf{Z}_{n}^{(t)})\mathbf{Tr}(\widetilde{\mathbf{U}}_{J,k}\mathbf{Z}_{n})\right\}\\
			=&\text{Re}\left\{ \mathbf{Tr}\left(\sum_{k\in\bm{\Omega}_{J}}C_{\mathbf{w}_{n},\mathbf{x}_{n,J}}^{(t)}(-k)\widetilde{\mathbf{U}}_{J,k}\mathbf{Z}_{n}\right) \right\}\\
			=&\text{Re}\left\{\mathbf{Tr}(\mathbf{Q}_{2}\mathbf{Z}_{n})\right\},
		\end{split}
	\end{equation}
	where 
	\begin{equation}\label{key}
		\mathbf{Q}_{2}
		=\begin{bmatrix}
			\mathbf{O}&\mathbf{J}^{H}\bm{\Phi}_{J}\\
			\mathbf{O}&\mathbf{O}
		\end{bmatrix},\;\bm{\Phi}_{J}=\sum\limits_{k\in\bm{\Omega}_{J}}C_{\mathbf{w}_{n},\mathbf{x}_{n,J}}^{(t)}(-k)\mathbf{U}_{k}.
	\end{equation}\par
	Since $ \text{vec}(\mathbf{Z}_{n})^{H}(\lambda_{u}\mathbf{I})\text{vec}(\mathbf{Z}_{n}^{(t)})=\lambda_{u}\mathbf{Tr}(\mathbf{Z}_{n}^{(t)}\mathbf{Z}_{n})$, (64) can be written as
	\begin{equation}\label{key}
		\begin{split}
			&\text{Re}\left\{\text{vec}(\mathbf{Z}_{n})^{H}(\mathbf{W}-\lambda_{u}\mathbf{I})\text{vec}(\mathbf{Z}_{n}^{(t)})\right\}\\
			=&\text{Re}\left\{\rho\mathbf{Tr}(\mathbf{Q}_{1}\mathbf{Z}_{n})+(1-\rho)\mathbf{Tr}(\mathbf{Q}_{2}\mathbf{Z}_{n})-\lambda_{u}\mathbf{Tr}(\mathbf{Z}_{n}^{(t)}\mathbf{Z}_{n})\right\}\\
			=&\text{Re}\left\{\mathbf{z}_{n}^{H}\left(\rho\mathbf{Q}_{1}+(1-\rho)\mathbf{Q}_{2}-\lambda_{u}\mathbf{z}_{n}^{(t)}(\mathbf{z}_{n}^{(t)})^{H}\right)\mathbf{z}_{n}\right\}\\
			=&\text{Re}\left\{\mathbf{z}_{n}^{H}(\mathbf{Q}-\lambda_{u}\mathbf{z}_{n}^{(t)}(\mathbf{z}_{n}^{(t)})^{H})\mathbf{z}_{n}\right\},
		\end{split}
	\end{equation}
	where $\mathbf{Q}=\rho\mathbf{Q}_{1}+(1-\rho)\mathbf{Q}_{2} $. The proof is completed. $ \hfill\blacksquare $\par
	%\section*{Acknowledgment}

	\ifCLASSOPTIONcaptionsoff
	\newpage
	\fi

	%\begin{figure*}[!t]
	%\centering
	%\subfloat[Case I]{\includegraphics[width=2.5in]{box}%
	%\hfil
	%\subfloat[Case II]{\includegraphics[width=2.5in]{box}%
	%\label{fig_second_case}}
	%\caption{Simulation results for the network.}
	%\label{fig_sim}
	%\end{figure*}

	
	%\end{IEEEbiography}
	
	%	Biography text here.
	
	%\newpage
	
	
	
	%\vfill
	


\begin{thebibliography}{1}
		
		\bibitem{L. Zheng-2019}
		L. Zheng, M. Lops, Y. C. Eldar, and X. Wang, ``Radar and communication coexistence: An overview: A review of recent methods,'' \emph{IEEE Signal Process. Mag.}, vol. 36, no. 5, pp. 85–99, Sep. 2019.
		
		%\bibitem{F. Liu-2020(1)}
		%F. Liu, C. Masouros, A. P. Petropulu, H. Griffiths, and L. Hanzo, ``Joint radar and communication design: Applications, state-of-the-art, and the
		%road ahead,'' \emph{IEEE Trans. Commun.}, vol. 68, no. 6, pp. 3834–3862, 2020.
		\bibitem{S. Lu-2024}
		S. Lu, F. Liu, Y. Li, K. Zhang, H. Huang et al.,  ``Integrated sensing and communications: Recent advances and ten open challenges,'' \emph{IEEE Internet of Things Journal}, vol. 11, no. 11, pp. 19094-19120, 2024.
		
		\bibitem{M. Liu-2023}
		M. Liu, M. Yang, Z. Zhang, H. Li, F. Liu, A. Nallanathan, and L. Hanzo,  ``Sensing-communication coexistence vs. Integration,'' \emph{IEEE Trans. Veh. Technol.}, vol. 72, no. 6, pp. 8158-8163, 2023. 
		
		\bibitem{M. Bica-2019}
		M. Bica and V. Koivunen, ``Radar waveform optimization for target parameter estimation in 
		cooperative radar-communications systems,'' \emph{IEEE Trans. Aerosp. Electron. Syst.}, vol. 55, no. 5, pp. 2314–2326, Oct. 2019.\par
		
		\bibitem{B. Li-2017}
		B. Li and A. P. Petropulu, ``Joint transmit designs for coexistence of
		MIMO wireless communications and sparse sensing radars in clutter,'' \emph{IEEE Trans. Aerosp. Electron. Syst.}, vol. 53, no. 6, pp. 2846–2864, Dec. 2017.
		
		\bibitem{F. Liu-2018-1}
		F. Liu, C. Masouros, A. Li, T. Ratnarajah, and J. Zhou, ``MIMO radar and
		cellular coexistence: A power-efficient approach enabled by interference exploitation,'' \emph{IEEE Trans. Signal Process.}, vol. 66, no. 14, pp. 3681–3695, Jul. 2018.\par
		
		\bibitem{F. Liu-2022}
		F. Liu, Y. Cui, C. Masouros, J. Xu, T. X. Han, Y. C. Eldar, and S. Buzzi,
		``Integrated sensing and communications: Towards dual-functional wireless networks for 6G and beyond,'' \emph{IEEE J. Sel. Areas commun.}, vol. 40, no. 6, pp. 1728-1767, 2022.
		
		\bibitem{Y. Cui-2021}
		Y. Cui, F. Liu, X. Jing, and J. Mu, ``Integrating sensing and communications for ubiquitous IoT: Applications, trends, and challenges,'' \emph{IEEE Netw.}, vol. 35, no. 5, pp. 158–167, Sep. 2021.
		
		\bibitem{K. Meng-2023}
		K. Meng, Q. Wu, S. Ma, W. Chen, K. Wang, and J. Li, ``Throughput maximization for UAV-enabled integrated periodic sensing and communication,'' \emph{IEEE Trans. Wireless Commun.}, vol. 22, no. 1, pp. 671-687, Jan. 2023. 
		
		\bibitem{F. Dong-2023-resource}
		F. Dong, F. Liu, Y. Cui, W. Wang, K. Han, and Z. Wang, ``Sensing as a service in 6G perceptive networks: A unified framework for ISAC resource allocation,'' \emph{IEEE Trans. Wireless Commun.}, vol. 22, no. 5, pp. 3522-3536, 2023.
		
		\bibitem{X. X. Yu-2022}
		X. Yu, X. Yao, J. Yang, L. Zhang, L. Kong, and G. Cui, ``Integrated waveform design for MIMO radar and communication via spatio-spectral modulation,'' \emph{IEEE Trans. Signal Process.}, vol. 70, pp. 2293-2305, 2022.\par
		
		
		\bibitem{G. N. Saddik-2007}
		G. N. Saddik, R. S. Singh, and E. R. Brown, ``Ultra-wideband multifunctional communications/radar system,'' \emph{IEEE Trans. Micr. Theory and Techn.}, vol. 55, no. 7, pp. 1431-1437, 2007.\par
		
		\bibitem{D. Gaglione-2018}
		D. Gaglione, C. Clemente, C. V. Ilioudis, A. R. Persico, I. K. Proudler, J. J. Soraghan, and A. Farina, ``Waveform design for communicating radar systems using fractional Fourier transform,'' \emph{Digital Signal Process.},
		vol. 80, pp. 57-69, 2018. 
		
		
		\bibitem{J. Johnston-2022}
		J. Johnston, L. Venturino, E. Grossi, M. Lops, and X. Wang, ``MIMO OFDM dual-function radar-communication under error rate and beampattern constraints,'' \emph{IEEE J. Sel. Areas Commun.}, vol. 40, no. 6, pp. 1951-1964, 2022.\par
		
		\bibitem{M. F. Keskin-2021}
		M. F. Keskin, H. Wymeersch, and V. Koivunen, ``MIMO-OFDM joint radar-communications: Is ICI friend or foe?'' \emph{IEEE J. Sel. Topics Signal Process.}, vol. 15, no. 6, pp. 1393-1408, 2021.
		
		\bibitem{J. Yang-2020}
		J. Yang, G. Cui, X. Yu, and L. Kong, ``Dual-use signal design for radar and communication via ambiguity function sidelobe control,'' \emph{IEEE Trans. Veh. Technol.}, vol. 69, no. 9, pp. 9781–9794, Sep. 2020.
		
		\bibitem{S. Zhou-2019}
		S. Zhou, X. Liang, Y. Yu, and H. Liu, ``Joint radar-communications couse waveform design using optimized phase perturbation,'' \emph{IEEE Trans. Aerosp. Electron. Syst.}, vol. 55, no. 3, pp. 1227–1240, Jun. 2019.
		
		\bibitem{Y. Dong-2023-letter}
		Y. Dong, F. Liu, and Y. Xiong, ``Joint receiver design for integrated sensing and communications,'' \emph{ IEEE Commun. Lett.}, vol. 27, no. 7, pp. 1854-1858, 2023.
		
		\bibitem{A. Hassanien-2016}
		A. Hassanien, M. G. Amin, Y. D. Zhang, and F. Ahmad, ``Signaling strategies for dual-function radar communications: An overview,'' \emph{IEEE Aerosp. Electron. Syst. Mag.}, vol. 31, no. 10, pp. 36–45, Oct. 2016.
		
		\bibitem{C. Sahin-2017}
		C. Sahin, J. Jakabosky, P. M. McCormick, J. G. Metcalf, and S. D. Blunt,
		``A novel approach for embedding communication symbols into physical radar waveforms,''  \emph{Proc. IEEE Radar Conference (RadarConf)}, 2017, pp. 1498–1503.
		
		\bibitem{D. Ma-2020}
		D. Ma, N. Shlezinger, T. Huang, Y. Liu, and Y. C. Eldar, ``Automotive dual-function radar communications systems: An overview,'' \emph{Proc. IEEE 11th Sens. Array Multichannel Signal Process. Workshop (SAM)},
		2020, pp. 1–5.
		
		\bibitem{C. Sturm-2011}
		C. Sturm and W. Wiesbeck, ``Waveform design and signal processing aspects for fusion of wireless communications and radar sensing,''
		\emph{Proceedings of the IEEE}, vol. 99, no. 7, pp. 1236–1259, 2011.\par
		
		\bibitem{Y. Liu-2017}
		Y. Liu, G. Liao, Z. Yang, and J. Xu, ``Multiobjective optimal waveform design for OFDM integrated radar and communication systems,'' \emph{Signal
			Process.}, vol. 141, pp. 331–342, 2017.\par
		
		\bibitem{Y. Huang-2022}
		Y. Huang, S. Hu, S. Ma, Z. Liu, and M. Xiao, ``Designing low-PAPR waveform for OFDM-based 
		RadCom systems,'' \emph{IEEE Trans. Wireless Commun.}, vol. 21, no. 9, pp. 6979-6993, 
		Sep. 2022. 
		
		
		
		\bibitem{F. Liu-2018-2}
		F. Liu, L. Zhou, C. Masouros, A. Li, W. Luo, and A. Petropulu, ``Toward dual-functional radar-communication systems: optimal waveform design,'' \emph{IEEE Trans. Signal Process.}, vol. 66, no. 16, pp. 4264-4279,  Aug. 2018.\par
		
		\bibitem{F. Liu-2018-3}
		F. Liu, C. Masouros, A. Li, H. Sun, and L. Hanzo, ``MU-MIMO communications with MIMO radar: From co-existence to joint transmission,'' \emph{IEEE Trans. Wireless Commun.}, vol. 17, no. 4, pp. 2755–2770, Apr. 2018.\par
		
		
		\bibitem{X. Liu-2020}
		X. Liu, T. Huang, N. Shlezinger, Y. Liu, J. Zhou, and Y. C. Eldar, ``Joint transmit beamforming for multiuser MIMO communications and MIMO radar,” \emph{IEEE Trans. Signal Process.}, vol. 68, pp. 3929–3944, Aug. 2020.\par
		
		\bibitem{F. Dong-2023}
		F. Dong, W. Wang, X. Li, F. Liu, S. Chen, and L. Hanzo, ``Joint beamforming design for dual-functional MIMO radar and communication systems guaranteeing physical layer security," \emph{IEEE Trans. Green Commun. Netw.}, vol. 7, no. 1, pp. 537-549, 2023.
		
		\bibitem{L. Chen-2022}
		L. Chen, Z. Q. Wang, Y. Du, Y. Chen, and F. R. Yu, ``Generalized transceiver beamforming for DFRC with MIMO 
		radar and MU-MIMO communication,'' \emph{IEEE J. Sel. Areas  Commun.}, vol. 40, no. 6, pp. 1795 -1808, 2022.
		
		\bibitem{C. G. Tsinos-2021}
		C. G. Tsinos, A. Arora, S. Chatzinotas, and B. Ottersten,  ``Joint transmit waveform and receive filter design for dual-function radar-communication systems,'' \emph{IEEE J. Sel. Topics Signal Process.}, vol. 15, no. 6, pp. 1378-1392, Nov. 2021.\par
		
		
		
		
		
		\bibitem{S.D. Berger-2003}
		S.D. Berger, ``Digital radio frequency memory linear range gate stealer spectrum,''
		\emph{IEEE Trans. Aerosp. Electron. Syst.}, vol. 39, no. 2, pp. 725–735, Apr. 2003.\par
		
		\bibitem{S. Hanbali-2019}
		S. Hanbali, ``Technique to counter improved active echo cancellation based on ISRJ with frequency shifting,'' \emph{IEEE Sensors J.}, vol. 19, no. 20, pp. 9194-9199, 2019.
		
		
		\bibitem{W. Xiong-2017}
		W. Xiong, G. Zhang, and W. Liu, ``Efficient filter design against interrupted sampling repeater 
		jamming for wideband radar,'' \emph{EURASIP J. Adv. Signal Process.}, vol. 2017, no. 1, pp. 1-12, 2017.\par
		\bibitem{J. Chen-2019}
		J. Chen, W. Wu, S. Xu, Z. Chen, and J. Zou, ``Band pass filter design against interrupted sampling repeater jamming based on time-frequency analysis,'' \emph{IET Radar, Sonar\&Nav.}, vol 13, no, 10, pp. 1646-1654, 2019.\par
		
		\bibitem{S. Lu-2018}
		S. Lu, G. Cui, X. Yu, L. Kong, and X. Yang, ``Cognitive radar waveform design against signal-dependent modulated jamming,'' \emph{Progress In Electrom. Research}, vol. 80, pp. 59-77, 2018.\par
		
		\bibitem{K. Zhou-2020}
		K. Zhou, D. Li, Y. Su, and T. Liu, ``Joint design of transmit waveform and mismatch
		filter in the presence of interrupted sampling repeater jamming,'' \emph{IEEE Signal Process. Lett.}, vol. 27, pp. 1610–1614, 2020.\par
		
		\bibitem{K. Zhou-2022}
		K. Zhou, F. He, and Y. Su, ``Fast algorithm for joint waveform and filter design against
		interrupted sampling repeater jamming,'' \emph{J. Radar}, vol. 11, no. 2, pp. 264–277, 2022. \par
		
		\bibitem{M. Ge-2021}
		M. Ge, X. Yu, Z. Yan, G. Cui, and L. Kong, ``Joint cognitive optimization of transmit waveform and receive 
		filter against deceptive interference,'' \emph{Signal Process.}, vol. 185, pp. 1-15, 2021.\par
		
		%\bibitem{K. Zhou-SAR-ISRJ}
		%K. Zhou, D. Li, S. Quan, T. Liu, Y. Su, and F. He, ``SAR waveform
		%and mismatched filter design for countering interrupted-sampling repeater jamming,'' \emph{IEEE Trans. Geosci. Remote Sens.}, vol. 60, pp. 1-14, 2022.
		
		\bibitem{S. Wei-2024}
		S. Wei, Y. Fang, Y. Song, S. Wei, and L. Zhang, ``Joint jam perception and adaptive waveform optimization for anti-interrupted sampling repeater jamming," \emph{IEEE Trans. Aerosp. Electron. Syst.}, vol. 60, no. 1, pp. 1129-1147, Feb. 2024.\par
		
		%\bibitem{F. L. Wang-2022}
		%F. Wang, C. Peng, J. Yin, N. Li, Y. Li, and X. Wang, ``Joint design of Doppler-tolerant
		%complementary sequences and receiving filters against interrupted sampling repeater jamming,'' \emph{J.
		%	Radar}, vol. 11, no. 2, pp. 278–288, 2022. \par
		
		\bibitem{Y. Zhou-2023}
		Y. Zhou, Q. Shi, Z. Zhou, and Y. Yang, ``Waveform design for integrated sensing and communication in the presence of DRFM forwarding interference,'' \emph{IET International Radar Conference (IRC 2023),} Chongqing, China, 2023, pp. 1316-1321.
		
		\bibitem{J. Li-2006}
		J. Li, J. R. Guerci, and L. Xu, ``Signal waveform's optimal under restriction design for active sensing,'' \emph{Fourth IEEE Workshop on Sensor Array and Multichannel Processing}, 2006, pp. 382-386.
		
		
		\bibitem{J. H. Wang-2021}
		J. Wang, P. Fan, Z. Zhou, and Y. Yang, ``Quasi-orthogonal Z-complementary pairs and their applications in fully polarimetric radar systems,'' \emph{IEEE Trans. Inf. Theory}, vol. 67, no. 7, pp. 4876-4890, 2021.\par
		
		
		%\bibitem{S. Lu-2016}
		%S. Lu, W. Yi, G. Cui, L. Kong, and X. Yang, “Design and application of dynamic environmental knowledge base,”
		%\emph{IET Radar, Sonar $\&$ Nav.}, vol. 10, no. 6, pp. 1118–1126, 2016.
		
		\bibitem{C. Zhou-2018}
		C. Zhou, Q. Liu, and X. Chen, ``Parameter estimation and suppression for DRFM‐based interrupted sampling repeater jammer,'' \emph{IET Radar, Sonar $\&$ Nav.}, vol. 12, no. 1, pp. 56-63, 2018.
		
		\bibitem{F. L. Wang-2021}
		F. Wang, C. Peng, J. Zhou, Y. Li, and X. Wang, ``Design of complete complementary sequences for 
		ambiguity functions optimization with a PAR constraint,'' \emph{IEEE Geosci. Remote Sens. 
			Lett.}, vol 19, pp. 1-5, 2021.\par
		
		\bibitem{M. Soltanalian-2014}
		M. Soltanalian and P. Stoica, ``Designing unimodular codes via quadratic optimization,'' \emph{IEEE Trans. Signal Process.}, vol. 62, no. 5, pp. 1221–1234, Mar. 2014.\par
		
		\bibitem{Y. Chen-2024}
		Y. Chen, Z. Chen, Y. Zhang, J. Yang, and D. Li, ``Joint design of doppler resilient unimodular discrete phase sequence waveform and receiving filter for multichannel radar,'' \emph{IEEE Trans. Signal Process.}, 2024.
		
		\bibitem{J. Song-2015}
		J. Song, P. Babu, and D. P. Palomar, “Optimization methods for designing sequences with low autocorrelation sidelobes," \emph{IEEE Trans. Signal Process.}, vol. 63, no. 15, pp. 3998-4009, Aug. 2015.\par
		
		\bibitem{J. Song-2016}
		J. Song, P. Babu, and D. P. Palomar, ``Sequence design to minimize the weighted integrated and peak sidelobe levels,'' \emph{IEEE Trans. Signal Process.}, vol. 64, no. 8, pp. 2051-2064, Apr. 2016. \par
		
		
		
		%\bibitem{J. Bolte-2014}
		%J. Bolte, S. Sabach, and M. Teboulle, ``Proximal alternating linearized minimization for nonconvex and nonsmooth problems,'' \emph{Math. Program.},
		%vol. 146, pp. 459–494, Aug. 2014.
		
		\bibitem{J. Geiping-2018}
		J. Geiping and M. Moeller, ``Composite optimization by nonconvex majorization-minimization,'' \emph{SIAM J. Imag. Sci.}, vol. 11, no. 4, pp. 2494–2528, Jan. 2018.
		
		
		\bibitem{K. Zhou-2022-J}
		K. Zhou, S. Quan, D. Li, T. Liu, F. He, and Y. Su, ``Waveform and filter joint design method for pulse compression sidelobe reduction,'' \emph{IEEE Trans. Geosci. Remote Sens.}, vol. 60, pp 1-15, 2022.\par
		
		\bibitem{H. B. Chang-2019}
		H. B. Chang, P. Enfedaque, and S. Marchesini, ``Blind  ptychographic  phase  retrieval  via
		convergent alternating  direction  method  of  multipliers,'' \emph{SIAM J. Imag. Sci.}, vol. 12, no. 1, pp. 153–185, 2019.\par
		
		%\bibitem{J. H. Wang-2022}
		%J. Wang, P. Fan, Q. Shi, and Z. Zhou, ``Doppler resilient integrated sensing and communication waveforms design,'' \emph{J. Radar}, vol. 11, pp. 1-12, 2022.\par
		
		\bibitem{J. Wu-2015}
		J. Wu, G. Cui, and L. Kong, ``A cognitive waveform design approach against velocity deception jamming,'' \emph{Radar Sci. Technol.}, vol. 13, no. 2, pp. 133-138, 2015.\par 
		
		
		\bibitem{LPG}
		O. Rabaste and L. Savy, ``Mismatched filter optimization for radar applications using quadratically constrained quadratic programs,'' \emph{IEEE Trans. Aerosp. Electron. Syst.}, vol. 51, no. 4, pp. 3107–3122, Oct. 2015.
		
		\bibitem{M. A. Richards-2014}
		M. A. Richards, \emph{Fundamentals of Radar Signal Processing}. New York,
		NY, USA: McGraw-Hill Educ., 2014.
		
		%\bibitem{Dingyou-2022}
		% D. Y. Ma, X. Liu, T. Y. Huang, et al., ``Joint radar and communications: Shared waveform designs and performance bounds,'' \emph{Journal of Radars}, 2022, 11(2): 198-212.\par
	\end{thebibliography}
\end{document}